\documentclass{aa}
\usepackage{newtxtext,newtxmath}
\usepackage[utf8]{inputenc}
\usepackage{ae,aecompl}
\usepackage{graphicx}
\usepackage{amsmath,amssymb}
\usepackage{subfig}
\usepackage{natbib}
\usepackage[breaklinks,colorlinks,citecolor=blue]{hyperref}
\usepackage{siunitx}
\usepackage[dvipsnames]{xcolor}
\usepackage{ulem}
\usepackage{placeins}
\usepackage[flushleft]{threeparttable}

\begin{document}

\title{Misalignment of the outer disk of DK\,Tau \\ and a first look at its magnetic field using spectropolarimetry}

\author{M.\,Nelissen\inst{1} \and P.\,McGinnis\inst{1} \and C.\,P.\,Folsom\inst{2,3} \and T.\,Ray\inst{1} \and A.\,A.\,Vidotto\inst{4} \and E.\,Alecian\inst{5} \and J.\,Bouvier\inst{5} \and J.\,Morin\inst{6} \and J.-F.\,Donati\inst{7} \and R.\,Devaraj\inst{1}}

\institute{ 
Dublin Institute for Advanced Studies, Astronomy \& Astrophysics Section, 
31 Fitzwilliam Place, Dublin\,2, Ireland \\
\email{nelissen@cp.dias.ie} 
\and Tartu Observatory, University of Tartu, Observatooriumi 1, Tõravere, 61602 Tartumaa, Estonia
\and University of Western Ontario, Department of Physics \& Astronomy, London, Ontario, N6A 3K7, Canada
\and Leiden Observatory, Leiden University, PO Box 9513, 2300RA, Leiden, The Netherlands
\and Univ. Grenoble Alpes, CNRS, IPAG, 38000 Grenoble, France
\and LUPM, Université de Montpellier \& CNRS, Montpellier, Cedex 05, France
\and Univ. de Toulouse, CNRS, IRAP, 14 avenue Belin, 31400 Toulouse, France
}

\date{Accepted 22 December 2022}

\abstract{
    {\textit{Context.} Misalignments between a forming star's rotation axis and its outer disk axis, although not predicted by standard theories of stellar formation, have been observed in several classical T\,Tauri stars (cTTs). The low-mass cTTs DK\,Tau is suspected of being among them. In addition, it is an excellent subject to investigate the interaction between stellar magnetic fields and material accreting from the circumstellar disk, as it presents clear signatures of accretion. }
    
    {\textit{Aims.} The goal of this paper is to study DK\,Tau's average line-of-sight magnetic field in both photospheric absorption lines and emission lines linked to accretion, using spectropolarimetric observations, as well as to examine inconsistencies regarding its rotation axis.}
    
    {\textit{Methods.} We used data collected with the ESPaDOnS spectropolarimeter, at the Canada-France-Hawaii Telescope, and the NARVAL spectropolarimeter, at the Télescope Bernard Lyot, probing two distinct epochs (December 2010 to January 2011 and November to December 2012), each set spanning a few stellar rotation cycles. 
We first determined the stellar parameters of DK\,Tau, such as effective temperature and $v\sin i$. Next, we removed the effect of veiling from the spectra, then obtained least-squares deconvolution (LSD) profiles of the photospheric absorption lines for each observation, before determining the average line-of-sight magnetic field from them.
We also investigated accretion-powered emission lines, namely the 587.6\,nm He{\sc i} line and the Ca{\sc ii} infrared triplet (at 849.8\,nm, 854.2\,nm and 866.2\,nm), as tracers of the magnetic fields present in the accretion shocks. }
    
    {\textit{Results}. We find that DK\,Tau experiences accretion onto a magnetic pole at an angle of $\sim30\si{\degree}$ from the pole of its rotation axis, with a positive field at the base of the accretion funnels. In 2010 we find a magnetic field of up to 0.95kG (from the Ca{\sc ii} infrared triplet) and 1.77kG (from the He{\sc i} line) and in 2012 we find up to 1.15kG (from the Ca{\sc ii} infrared triplet) and 1.99kG (from the He{\sc i} line). 
    Additionally, using our derived values of period, $v \sin i$ and stellar radius, we find a value of 58\si{\degree} (+18)(-11) for the inclination of the stellar rotation axis, which is significantly different from the outer disk axis inclination of 21\si{\degree} given in the literature.}
    
    {\textit{Conclusion}. We find that DK\,Tau's outer disk axis is likely misaligned compared to its rotation axis by 37\si{\degree}.}
}

\keywords{Stars: individual: DK\,Tau - Stars: variables: T Tauri - Stars: magnetic field - Accretion, accretion disks - Techniques: polarimetric - Techniques: spectroscopic}

\titlerunning{Misalignment of the outer disk of DK\,Tau}
\maketitle


\section{Introduction} \label{Intro}

Stellar magnetic fields are omnipresent and play an essential role in the formation of stars and planets. Understanding their impact is therefore crucial to the study of stellar birth. We do not, however, yet possess a complete picture of how stellar magnetic fields originate, how they evolve over time, or the extent of their impact on circumstellar disks and the accretion process in the early stages of a star's life \cite[see e.g.,][]{2007prpl.conf..479B, 2012ApJ...755...97G, 2016MNRAS.457..580F, 2016ARA&A..54..135H, 2019A&A...622A..72V}.
Stellar magnetic fields of accreting T\,Tauri stars play an essential role in driving accretion and strongly impact the geometry of the accretion flow \cite[see][]{2016ARA&A..54..135H}. 
By analyzing the magnetic field along the line-of-sight and integrated over the visible stellar hemisphere measured in the acc retion-powered emission lines, one can recreate a picture of the component of the stellar magnetic field that dominates the accretion process. 
This is an integral quantity that relates Stokes\,I and Stokes\,V, making use of spectropolarimetric data. 
The interested reader is referred to, for example, \cite{1979A&A....74....1R, 2009ARA&A..47..333D, 2012EAS....57..165M}. 
The Stokes parameters and the magnetic field are connected through the Zeeman effect. This effect describes the impact of a magnetic field on a spectrum: its atomic (and molecular) lines are broadened or split, depending on the strength of the field and the sensitivity of the line in question \cite[see e.g.,][]{2011asia.book.....T, 2014IAUS..302...25H}.

In this work, we analyze the spectropolarimetry of the classical T\,Tauri star (cTTs) DK\,Tau. DK\,Tau is a young low-mass star surrounded by a circumstellar disk which is actively accreting from its inner regions. 
It is a wide binary \cite[separation 2\farcs 38, equivalent to 307\,au - see e.g.,][]{2019A&A...628A..95M}, 
which allows DK\,Tau~A (hereafter "DK Tau") to be spatially resolved with spectropolarimetry and studied on its own. 
It is located in the Taurus Molecular Cloud at a distance of 132.6\,pc 
\cite[][]{gaia1, gaia2, 2022arXiv220800211G}. 
Its spectral type is K7 \cite[see e.g.,][]{2007ApJ...664..975J, 2011ApJ...730...73F} 
with a heliocentric radial velocity of +16.2 km\,s$^{-1}$ \citep{2019AJ....157..196K}. 
\cite{2022arXiv220103588R} measured the inclination, with respect to the line of sight, of its outer (>20\,au) gaseous disk axis via projection to be 21\si{\degree}, using the $^{\rm 12}$CO\,(J\,=\,2–1) line with ALMA. 
The star’s rotation period $P$, as determined using photometry, has been measured as 8.4\,days \cite[][]{1993AAS..101..485B} 
and 8.18\,days \cite[][]{2010PASP..122..753P, 2012AstL...38..783A}. 

DK\,Tau presents with significant and variable veiling \cite[a strong signature of accretion - see e.g.,][]{hartigan95, 2011ApJ...730...73F}. 
In addition to accretion, it also shows evidence of ejection \cite[see e.g.,][]{hartigan95}, 
in particular inner disk winds and a jet as revealed by the detection of both low and high velocity forbidden [O{\sc i}] emission by \cite{hartigan95} and \cite{2019ApJ...870...76B}.

The effect of veiling on spectra complicates magnetic field measurements, yet the study of active accretors is very valuable for understanding the interaction between stellar magnetic fields and accreting material from the circumstellar disk. 
Indeed, our current understanding of magnetospheric accretion \cite[see e.g.,][]{1994ApJ...429..781S, 2002ApJ...578..420R, 2008A&A...478..155B, 2016ARA&A..54..135H} 
involves the truncation of the inner circumstellar disk at a few stellar radii and the channeling of accretion in funnel flows by the stellar magnetic field. When the accreted matter falls onto the star at near free-fall velocities, it produces an accretion shock close to the stellar surface, which gives rise to continuum and line veiling \cite[][]{1998ApJ...509..802C} 
and generates a localized bright/hot spot at the level of the chromosphere. 

Simple models of stellar formation do not account for a misalignment between the star's rotation axis and its outer disk axis. 
However, misalignments may be common, as several dippers display a low inclination of their outer disk axis \cite[see e.g.,][]{2020MNRAS.492..572A, 2020A&A...633A..37S}. 
Dippers are stars that show flux dips in their light curves. The standard explanation involves circumstellar material from the inner disk passing in front of the star, occulting it periodically or aperiodically \cite[see e.g.,][]{2015A&A...577A..11M, 2021A&A...651A..44R}. 
Dippers therefore require a relatively high inclination for the inner disk axis (which is here assumed to point in the same direction as the stellar rotation axis), that is inconsistent with their measured outer disk axis inclination. 
Such a misalignment has been directly measured in few cTTs \cite[see e.g.,][]{2018A&A...620A.195A, 2020A&A...636A.108B}. 
These observations suggest a more complex formation mechanism than normally considered, though it is not clear what gives rise to such a misalignment. As examples of potential causes, \cite{2020A&A...633A..37S} suggest the possibility of two different protostellar collapses, whereas \cite{2018A&A...620A.195A} invoke the presence of a massive planet inside the disk gap, and \cite{2018A&A...619A.171B} the effects of a low-mass stellar companion. 
DK\,Tau shows signs of being one of these systems with a misaligned outer disk. 

In this paper, we describe our observations in Sect.\,\ref{Obs}. We detail our analysis and results regarding stellar parameters, veiling, and magnetic characterizations in Sect.\,\ref{Anres}. 
In Section\,\ref{Disc} we discuss the implications of the inclination we measure for DK\,Tau and its magnetic field. 
Finally, Sect.\,\ref{Concl} contains our conclusions.


\section{Observations} \label{Obs}

Our data set is comprised of two sets of circularly polarized spectra of DK\,Tau, collected from December 2010 to January 2011, and from the end of November to the end of December 2012, with the spectropolarimeters ESPaDOnS (Echelle SpectroPolarimetric Device for the Observation of Stars), mounted at the CFHT (Canada-France-Hawaii Telescope) 3.6 meter telescope in Hawaii \citep{2006ASPC..358..362D}, 
and NARVAL, mounted at the 2 meter TBL (Télescope Bernard Lyot) on the Pic du Midi in France \citep{2003EAS.....9..105A}. 
These echelle spectropolarimeters cover the visible domain, from 370 to 1\,050\,nm, in a single exposure, and have a resolving power of 65\,000. ESPaDOnS has a fiber aperture of 1\farcs 66, while NARVAL has one of 2\farcs 80.

Table\,\ref{TableDates} lists the dates of the middle of the observations for the two ESPaDOnS and NARVAL data sets. The total exposure time was 4\,996.0\,s for each ESPaDOnS observation and 4\,800.0\,s for each NARVAL observation. 
In 2010 a total of 15 observations were taken over 39 days, and in 2012 a total of 12 observations were taken over 35 days, with the intention of capturing a few rotation cycles of the target in each set. 

The data are public and were downloaded from the archive of the PolarBase website\footnote{\url{http://polarbase.irap.omp.eu}} \cite[see e.g.,][]{2014PASP..126..469P}. These observations were made as a result of proposals 10BP12 and 12BP12, with J.-F.\,Donati as P.I in both cases and obtained as part of the MaPP (Magnetic Protostars and Planets) large program at the CFHT. 
We also downloaded the corresponding image files for the ESPaDOnS data from the Canadian Astronomy Data Centre (CADC) website\footnote{\url{http://www.cadc-ccda.hia-iha.nrc-cnrc.gc.ca/en}}. 
The data had been previously reduced at the CFHT and TBL. The reduction was carried out with the LibreESpRIT (for "Echelle Spectra Reduction: an Interactive Tool") reduction package specifically built for extracting polarization echelle spectra from raw data. This includes subtracting the bias and the dark frames, and correcting for the variations in sensitivity using flat field frames \citep{1997MNRAS.291..658D}. 
The spectra were continuum normalized in addition to the LibreESpRIT automatic continuum normalization, as the automatic procedure is not tailored for stars presenting with emission lines and it did not manage to properly adjust the continuum. 

\begin{table*}
\begin{center}
\caption{Dates for the 2010 and 2012 ESPaDOnS and NARVAL data sets. \label{TableDates}}
\begin{tabular}{c c c c c c}
\hline
 Date & Heliocentric Julian & Rotation cycle & S/N of the continuum & Airmass & Instrument \\
 (yyyy-mm-dd) & date (UTC) & (8.2 day period) & at the central wavelength & &  \\
\hline 
  2010-11-26 & 2\,455\,527.436\,03 & 0.00 & 70  & 1.1 & NARVAL \\  
  2010-12-09 & 2\,455\,540.393\,30 & 1.58 & 77  & 1.2 & NARVAL \\  
  2010-12-10 & 2\,455\,541.397\,11 & 1.70 & 53  & 1.1 & NARVAL \\  
  2010-12-13 & 2\,455\,544.418\,52 & 2.07 & 42  & 1.1 & NARVAL \\  
  2010-12-14 & 2\,455\,544.979\,74 & 2.14 & 72  & 1.1 & ESPaDOnS \\  
  2010-12-15 & 2\,455\,545.853\,82 & 2.25 & 97  & 1.0 & ESPaDOnS \\  
  2010-12-16 & 2\,455\,546.854\,61 & 2.37 & 100  & 1.0 & ESPaDOnS \\  
  2010-12-17 & 2\,455\,547.826\,04 & 2.49 & 108  & 1.1 & ESPaDOnS \\  
  2010-12-18 & 2\,455\,548.819\,03 & 2.61 & 113  & 1.1 & ESPaDOnS \\  
  2010-12-19 & 2\,455\,549.786\,41 & 2.73 & 81  & 1.2 & ESPaDOnS \\  
  2010-12-19 & 2\,455\,550.390\,86 & 2.80 & 54  & 1.1 & NARVAL \\  
  2010-12-24 & 2\,455\,554.883\,63 & 3.35 & 83  & 1.0 & ESPaDOnS \\  
  2010-12-26 & 2\,455\,557.034\,94 & 3.61 & 101  & 1.8 & ESPaDOnS \\  
  2010-12-30 & 2\,455\,560.977\,48 & 4.09 & 96  & 1.3 & ESPaDOnS \\  
  2011-01-03 & 2\,455\,565.451\,64 & 4.64 & 68  & 1.1 & NARVAL \\  
\hline
  2012-11-19 & 2\,456\,250.509\,43 & 0.00 & 68  & 1.1 & NARVAL \\  
  2012-11-25 & 2\,456\,256.919\,80 & 0.78 & 91  & 1.0 & ESPaDOnS \\  
  2012-11-28 & 2\,456\,259.895\,31 & 1.15 & 121  & 1.1 & ESPaDOnS \\  
  2012-11-29 & 2\,456\,260.991\,38 & 1.28 & 127  & 1.1 & ESPaDOnS \\  
  2012-12-01 & 2\,456\,262.947\,48 & 1.52 & 105  & 1.0 & ESPaDOnS \\  
  2012-12-02 & 2\,456\,263.865\,70 & 1.63 & 84  & 1.1 & ESPaDOnS \\  
  2012-12-04 & 2\,456\,265.965\,15 & 1.89 & 120  & 1.0 & ESPaDOnS \\  
  2012-12-07 & 2\,456\,268.846\,50 & 2.24 & 94  & 1.1 & ESPaDOnS \\  
  2012-12-09 & 2\,456\,271.387\,97 & 2.55 & 70  & 1.2 & NARVAL \\  
  2012-12-10 & 2\,456\,271.825\,04 & 2.60 & 107  & 1.2 & ESPaDOnS \\  
  2012-12-12 & 2\,456\,273.557\,58 & 2.81 & 63  & 1.2 & NARVAL \\  
  2012-12-23 & 2\,456\,284.762\,49 & 4.18 & 95  & 1.3 & ESPaDOnS \\  
\hline
\end{tabular}
\end{center}
\end{table*}


\section{Analysis \& results} \label{Anres}

\subsection{Veiling} \label{Veiling}

Accretion shocks are at a higher temperature than the photosphere. This adds an extra continuum to the stellar continuum, artificially decreasing the depth of the photospheric absorption lines. This is known as veiling and it varies with the wavelength, and can also vary from night to night. Its effect needs to be removed from the spectra in order to analyze the stellar magnetic field from the photospheric absorption lines.

Veiling ($R$) is defined as the ratio between the flux of the accretion shock and the photospheric flux.
For a normalized spectrum, it can be expressed using the following equation:
\begin{equation}
I_{\textrm{v}}(\lambda) = [I_{\textrm{ph}}(\lambda) + R(\lambda)] N(\lambda)
\label{EquationVeil}
\end{equation} 
where $I_{\textrm{v}}(\lambda)$ refers to the veiled intensity at wavelength $\lambda$, $I_{\textrm{ph}}(\lambda)$ to the intensity of the photosphere at wavelength $\lambda$, $R(\lambda)$ to the veiling at wavelength $\lambda$ and $N(\lambda)$ is a normalization constant at wavelength $\lambda$.

In order to measure the veiling in our spectra, we used a technique based on the fitting of a rotationally broadened and artificially veiled weak-lined T\,Tauri star (wTTs) spectrum to the spectrum of DK\,Tau. By choosing a wTTs with the same spectral type as our cTTs and coming from the same star forming region\footnote{This implies that both T\,Tauri stars would have the same chemical composition and very similar age and $\log g$. We also assume that the microturbulence and macroturbulence velocities should be very similar.}, this wTTs can be seen as the purely photospheric version of our star because it experiences no accretion. 
We tested several wTTs with a K7 spectral type and a line-of-sight-projected equatorial rotational velocity $v \sin i$ lower than the $v \sin i$ of DK\,Tau (see Sect.\,\ref{Param}), in order to find the one that provided the best fit. We ultimately used the spectrum of TAP45 \cite[which has a $v \sin i$ of 11.5\,km\,s$^{-1}$ - see][]{1987AJ.....94.1251F, 1993AAS..101..485B} as a template to estimate the veiling across DK\,Tau's spectra as the extra continuum that has to be added to the wTTs spectrum in order to reproduce the veiled cTTs spectrum (at the lowest $\chi^{2}$ level).

\begin{figure}[htb]
\begin{center}
\begin{minipage}{7.5cm}
\includegraphics[scale=0.47]{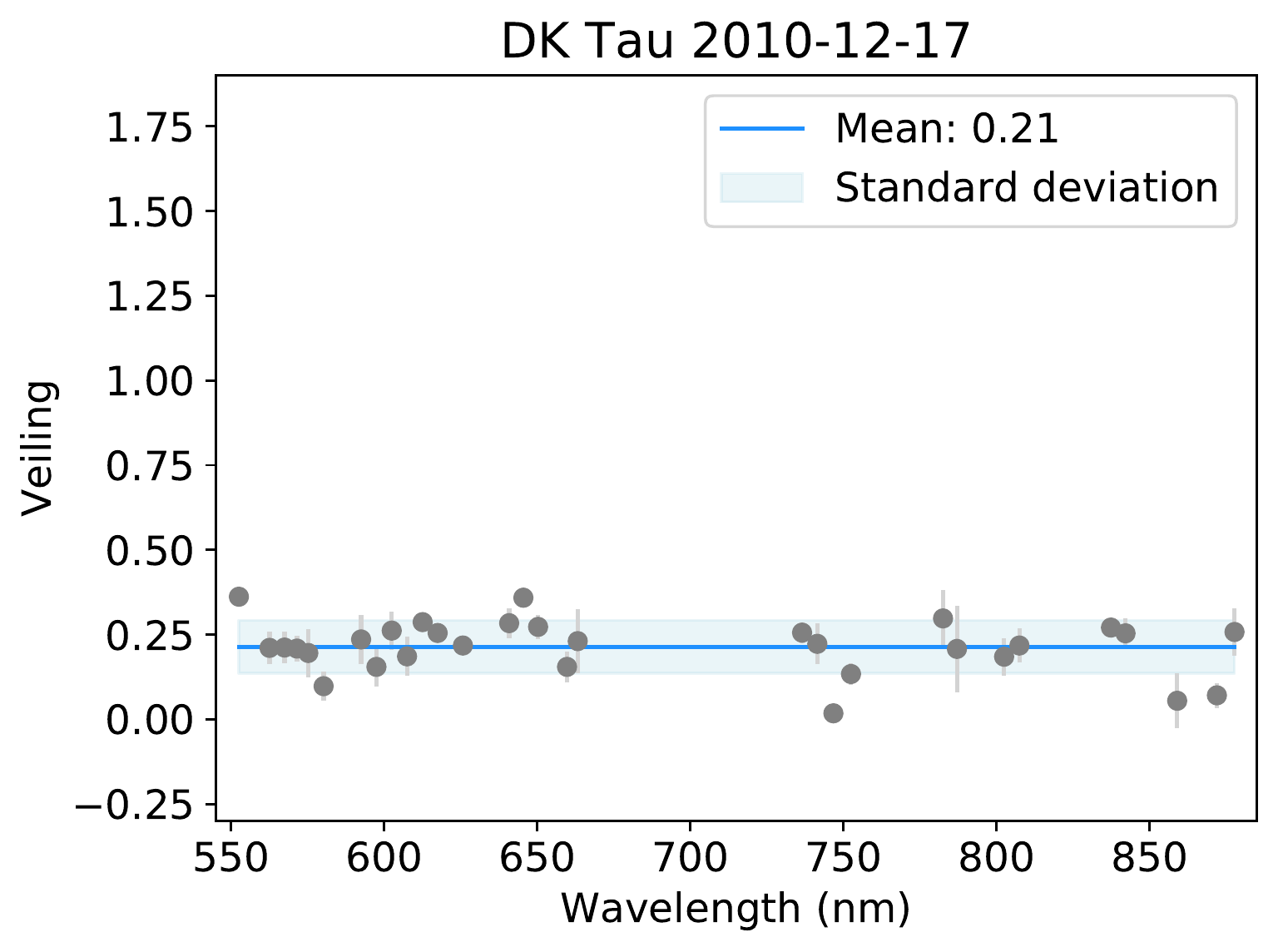}
\end{minipage}
\begin{minipage}{7.5cm}
\includegraphics[scale=0.47]{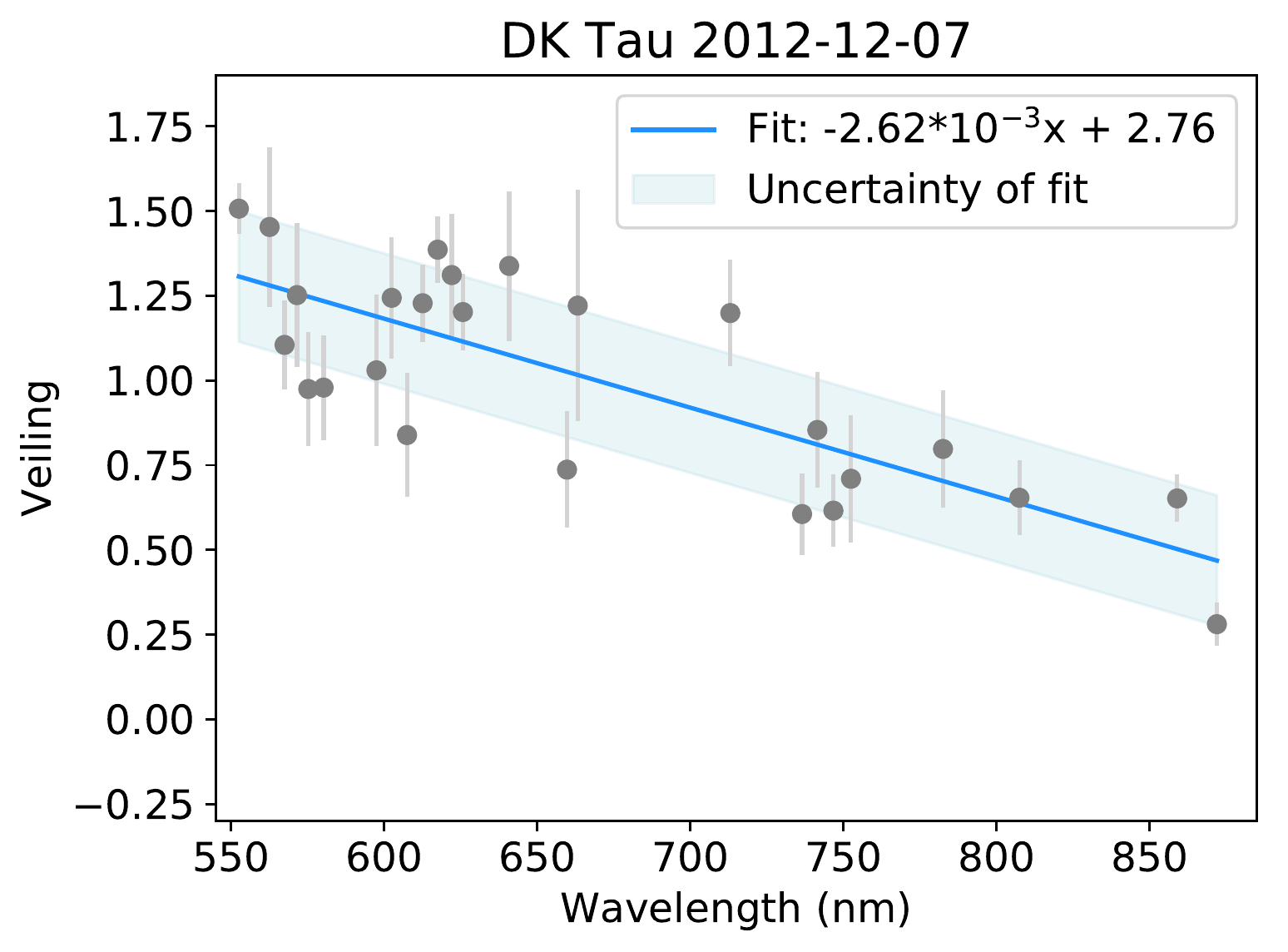}
\end{minipage}
\caption{Veiling (gray dots), the best linear fit (blue line) and the standard deviation (light blue shaded region) as a function of wavelength for the fourth night of the 2010 ESPaDOnS observations (top panel) and for the seventh night of the 2012 ESPaDOnS observations (bottom panel). \label{Veiling20102012}}
\end{center}
\end{figure}

We obtained values of the veiling as a function of the wavelength (in bins of $\sim$20\,nm), for each spectrum. When the value is less than $\sim$0.4 throughout the spectrum, we see that it is approximately constant in wavelength, therefore we took the mean value as $R$ for the whole spectrum.
When it is larger than this, we fit a linear relation through the points and considered this function as our $R(\lambda)$. 
We then inverted Eq.\,\ref{EquationVeil} to recover the normalized photospheric spectrum of each observation.
Figure\,\ref{Veiling20102012} shows the night with the least veiling and the one with the most veiling on the top and bottom panel respectively.

In 2010, the peak values of veiling (at $\sim$550\,nm) for each observation range from 0.2 to 1.8. Three observations out of fifteen have nearly constant veiling values across their spectrum. 
In 2012, the peak values of veiling for each observation range from 0.2 to 1.3 (lower than the value from two years before). Only one observation out of twelve has a nearly constant veiling value across its spectrum.

For the two sets of observations, we see that the slope of veiling as a function of wavelength steepens when the veiling is higher as well. We also find that when the veiling is high, there is more scatter. We believe this scatter is due to the correlation of stronger line emission with stronger mass accretion rate \cite[see e.g.,][]{2012AstL...38..649D, 2018A&A...610A..40R}. In that case, 
individual line emission would contribute more to the veiling by filling in absorption lines; 
whereas for nights when the veiling is low, continuum veiling would remain the dominant form of veiling. 
The interpretation of these trends goes beyond the scope of this work, and will be investigated in a subsequent paper. 


\subsection{Stellar parameters} \label{Param}

Given the importance of accurate stellar parameters, we took advantage of ESPaDOnS and NARVAL's continuum normalized high resolution spectra to derive precise values for DK\,Tau, 
in particular of the line-of-sight-projected equatorial rotational velocity $v \sin i$, needed to investigate the potential misalignment of DK\,Tau's outer disk, and for the effective temperature $T_{\textrm{eff}}$, which is needed for our analysis. 
For this, we first obtained the mean spectrum of the four nights with the least amount of veiling (i.e., $R$ = 0.2 for all four nights, see Sect. \ref{Veiling}) in order to get a spectrum with a higher signal to noise ratio of 136, whereas the individual spectra had a signal to noise ratio of 103 on average. 
We then used the ZEEMAN spectrum synthesis program, developed by \cite{1988ApJ...326..967L} and \cite{2001A&A...374..265W} and modified by \cite{2016MNRAS.457..580F} to derive the stellar parameters. Under the assumption of local thermodynamic equilibrium, this code solves radiative transfer equations. 
It iteratively compares a synthetic spectrum (calculated using a grid of model atmospheres and a list of atomic data) with the observation, by $\chi^{2}$ minimization, to determine free model parameters. Veiling was included in the model. 

We used the Vienna Atomic Line Database (VALD) website\footnote{\url{http://vald.astro.uu.se}} \cite[see e.g.,][]{2015PhyS...90e4005R} to acquire a continuous atomic line list ranging from 400\,nm to 1\,000\,nm for a star of $T_{\textrm{eff}}$ = 4\,000\,K, logarithmic surface gravity $\log g$ = 4 (in cgs units) and a solar metallicity. 
We used the MARCS model atmosphere grid of \cite{2008A&A...486..951G}. 
In the ZEEMAN code, we specified the following initial values for the stellar parameters:
for $T_{\textrm{eff}}$ and $v \sin i$, we used the values provided in the literature, of $T_{\textrm{eff}}$ = 4\,000\,K \cite[][]{2014ApJ...786...97H}, 
and $v \sin i$ = 12.7\,km\,s$^{-1}$ \cite[][]{2020MNRAS.497.2142M}. 
We used $\log g$ = 4 (in cgs units), as this is typical of T\,Tauri stars (TTs), 
microturbulence velocity $v_{\textrm{mic}}$ = 1\,km\,s$^{-1}$, macroturbulence velocity $v_{\textrm{mac}}$ = 0\,km\,s$^{-1}$, 
solar metallicity and a veiling of 0.1. 
We started with a model with initial values of $T_{\textrm{eff}}$, $v \sin i$, $\log g$ and veiling for a fixed value of  $v_{\textrm{mic}}$, $v_{\textrm{mac}}$ and metallicity. 
We then ran fits with only $T_{\textrm{eff}}$ and $v \sin i$ as free parameters.
When fitting the observation, we used a wavelength range from 400\,nm to 1\,000\,nm. We ran the code on several windows throughout the spectrum. We calculated the average of the values obtained for the different spectral windows and took the standard deviation of the spread as the error bars. 
We derive a $T_{\textrm{eff}}$ of 4\,150 $\pm$ 110 K and a $v \sin i$ of 13.0 $\pm$ 1.3 km\,s$^{-1}$. We note that our values are in good agreement with the ones found in the literature \cite[see][]{2014ApJ...786...97H, 2020MNRAS.497.2142M}. 

We also determined DK\,Tau's radius. 
We first calculated the stellar luminosity $L_\star$ using the J-band magnitudes of \cite{2007ApJ...669.1072E}. 
As DK\,Tau~A and B are not spatially resolved in their observations, we corrected the J-band magnitude to remove the contribution of DK\,Tau~B. This was done by extrapolating the brightness ratio given by \cite{2007ApJ...669.1072E} for the K-band (of 3.3) to the J-band, taking into consideration the shape of the continua of the two stars based on their respective spectral types, as well as the individual extinction each star suffers \citep[i.e., A$_{\textrm{V}}$ = 0.7\,mag for DK\,Tau~A, and A$_{\textrm{V}}$ = 1.80\,mag for DK\,Tau~B - ][]{2014ApJ...786...97H}. 
We find a flux ratio of 4.4 for the J-band. We then corrected the magnitude for the extinction in DK\,Tau~A (i.e., 0.7\,mag) \cite[][]{2014ApJ...786...97H}\footnote{We chose to use the value quoted by \cite{2014ApJ...786...97H} as they give a value for both components of the binary.}. 
It can be noted that the value for the extinction quoted by \cite{2011ApJ...730...73F} is a factor 2 higher. The variability of the extinction is probably connected to the dipper behavior of the star \citep{2021A&A...651A..44R}. 
We also corrected for a veiling of $R_J = 0.3 \pm 0.1$, based on an interpolation of our measurements of veiling as a function of wavelength (see Sect.\,\ref{Veiling}). This is very similar to the values observed by \cite{2011ApJ...730...73F}. 
We then used the bolometric correction in the J band from \cite{2013ApJS..208....9P}. 
We find a value of $L_\star$ = (1.65 $\pm$ 0.25) $L_{\odot}$. Using the relation $R^2 = L_\star / (4 \pi \sigma T_{\textrm{eff}}^4)$ and our measured value of $T_{\textrm{eff}}$, we derive a stellar radius of $R_\star$ = (2.48 $\pm$ 0.25) $R_{\odot}$. 

Table\,\ref{TableProp} summarizes the measured properties of DK\,Tau. The inclination of the outer disk axis was measured by \cite{2022arXiv220103588R} using ALMA observations. 
We derive a different value for the inclination $i$ of the stellar rotation axis (see Sect.\,\ref{ibeta}). 
The stellar rotation period $P$ mentioned in the literature is based on photometry \cite[see][]{1993AAS..101..485B, 2010PASP..122..753P, 2012AstL...38..783A}. The mass $M_\star$ was derived by \cite{2007ApJ...664..975J} from pre-main sequence evolutionary tracks. 
Using the \cite{2000A&A...358..593S} models\footnote{\url{http://www.astro.ulb.ac.be/~siess/pmwiki/pmwiki.php?n=WWWTools.PMS}}, we find that the values of  $T_{\textrm{eff}}$ and $L_\star$ that we obtain give a slightly higher $M_\star$ than the one of 0.7\,$M_\odot$ derived by \cite{2007ApJ...664..975J}, but it agrees with 0.7\,$M_\odot$ within 2$\sigma$ (see Appendix\,\ref{ParamSiess}). 

\begin{table}
\begin{center}
\begin{threeparttable}
\caption{Summary of the measured properties of DK\,Tau. \label{TableProp}}
\begin{tabular}{c c c}
\hline 
Stellar parameter & Value & Reference\\
\hline 
$i$ (\si{\degree}) & 58 (+18)(-11) & This work\\
Outer disk axis (\si{\degree}) & 21 $\pm$ 3 & \text{\cite{2022arXiv220103588R}}\\
$P$ (days) & 8.4 & \cite{1993AAS..101..485B}\\
 & 8.18 & \cite{2010PASP..122..753P}\\
 & 8.18 & \cite{2012AstL...38..783A}\\
 & 8.20 $\pm$ 0.13 & This work\\
$T_{\textrm{eff}}$ (K) & 4\,150 $\pm$ 110 & This work\\
$v \sin i$ (km\,s$^{-1}$) & 13.0 $\pm$ 1.3 & This work\\
$R_\star$ ($R_\odot$) & 2.48 $\pm$ 0.25 & This work\\
$M_\star$ ($M_\odot$) & 0.68 & \cite{2007ApJ...664..975J}\\
$L_\star$ ($L_{\odot}$) & 1.65 $\pm$ 0.25 & This work\\
\hline 
\end{tabular}
\begin{tablenotes}
\item The inclination is derived from $v \sin i = 2 \pi R_\star P^{-1} \sin i$ (see Sect. \ref{ibeta}).
\end{tablenotes}
\end{threeparttable}
\end{center}
\end{table}


\subsection{Least-squares deconvolution} \label{LSD}

Least-squares deconvolution (LSD) is a cross-correlation technique used to add the signatures from hundreds of photospheric absorption lines and obtain an average line profile of very high signal to noise ratio \cite[see e.g.,][]{1997MNRAS.291..658D, 2010A&A...524A...5K}. 
It makes use of a line mask, a list describing the position and depth of the chosen absorption lines. 
We first created a line mask using the VALD line list (with a detection threshold of 0.01, $T_{\textrm{eff}}$ = 4\,000\,K, $\log g$ = 4 and $v_{\textrm{mic}}$ = 1\,km\,s$^{-1}$) and assuming an ATLAS9 stellar atmosphere model \citep{1993KurCD..13.....K}. 
Next, using our line mask, we applied LSD to all observations 
in the range from 500\,nm to 1000\,nm (i.e., excluding the blue edge of the spectra because of excess noise), excluding regions with telluric and emission lines, 
as well as the lines most affected by accretion\footnote{We identified them by comparing with the spectrum of RU\,Lup, a cTTs of the same spectral type as DK\,Tau but presenting with more accretion, and determining the lines in emission.}, 
in order to obtain a mean line profile for each veiling-corrected spectrum. 
We set the effective Landé factor $\overline{g}_{0}$ = 1.4, the central intensity $d_{0}$ = 0.47 and the equivalent wavelength $\lambda_{0}$ = 650.0\,nm for all the nights \cite[see][]{2010A&A...524A...5K}. 
The LSD profiles were then normalized to be at the same equivalent width. The latter was done in order to correct for the residual effects of veiling.
For all nights, we obtained a definite Zeeman signal detection in the LSD profiles, with a false alarm probability smaller than 0.001\,\% (see the LSD profiles in Appendix\,\ref{LSDprof}).  

\begin{figure}[hbtp]
\begin{center}
\includegraphics[scale=0.4]{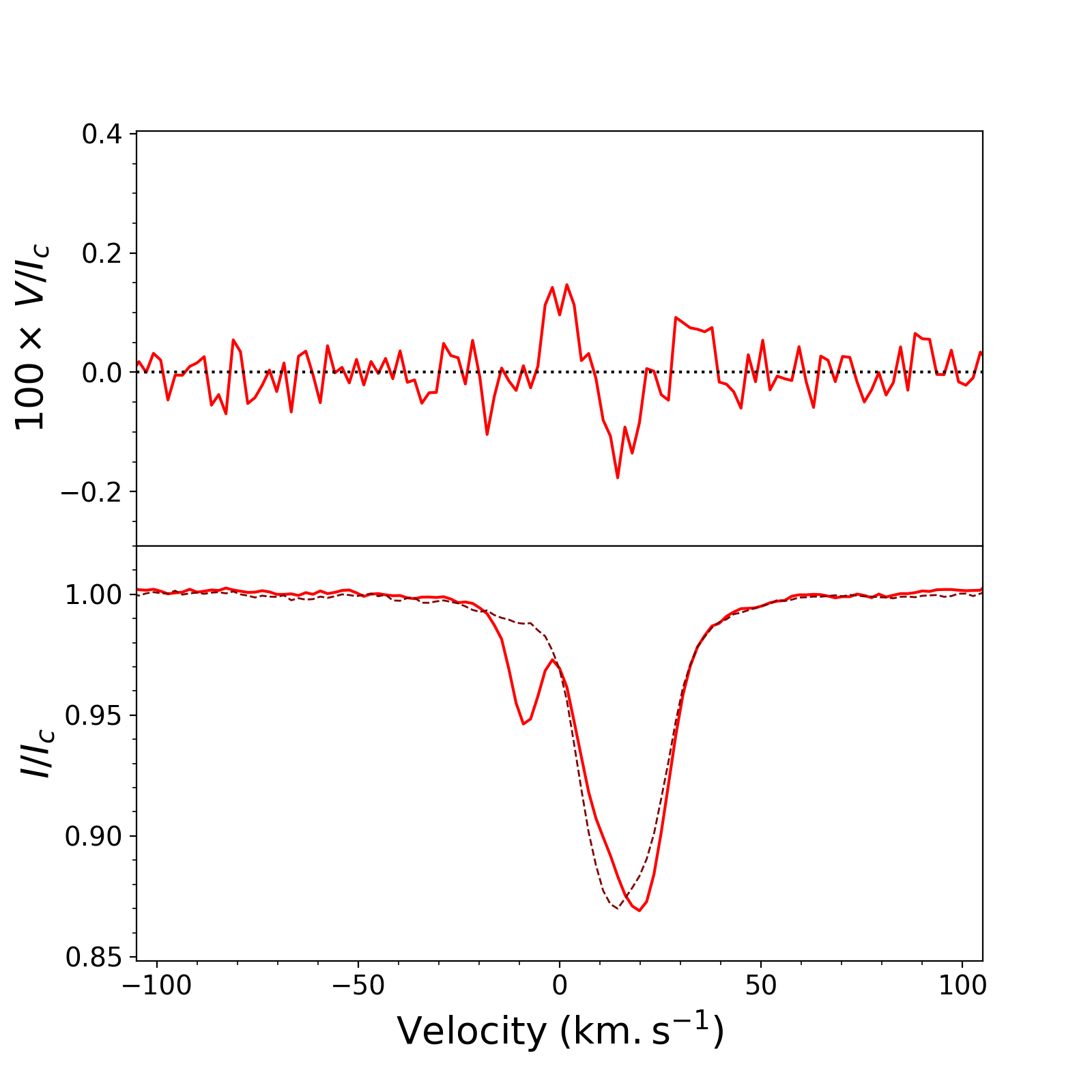}
\caption{LSD profile (in the heliocentric velocity frame) of Stokes\,V (top) and Stokes\,I (bottom) parameters normalized to the continuum for the sixth night of the 2010 ESPaDOnS observations. Note on this night scattered light from the Moon caused the small blue-ward absorption feature. For comparison, we also show the seventh night of the 2010 ESPaDOnS observations (dashed line), which did not suffer from moonlight contamination. \label{LSDBump}}
\end{center}
\end{figure}

The Stokes\,I LSD profiles of the ESPaDOnS and NARVAL observations made on 19 December 2010 presented a small absorption feature blueward of the main absorption line (see e.g., Fig.\,\ref{LSDBump}). The other nights all showed a single photospheric absorption line. 
This small absorption feature is due to scattered moonlight, as the full Moon was close to DK\,Tau on that night (with an angular separation of about 8\si{\degree}), and as the contamination is at the expected lunar radial velocity in the heliocentric rest frame\footnote{The online applet at \url{https://astroutils.astronomy.osu.edu/exofast/barycorr.html}, based on \cite{2014PASP..126..838W}, allows to calculate the correction applied to geocentric observations in order to transpose them in the heliocentric rest frame. In our case, this correction is -8.8\,km\,s$^{-1}$. Applied to the Moon's radial velocity of 0\,km\,s$^{-1}$ in the geocentric rest frame, this translates into a radial velocity of -8.8\,km\,s$^{-1}$ in the heliocentric rest frame.}. 
This type of contamination has been seen before \cite[see e.g.,][]{2011MNRAS.412.2454D}. 

We fit the wing of the main absorption feature with that of a Voigt profile and manually removed the contamination. 


\subsection{Average line-of-sight magnetic field} \label{Blos}

We measured the magnetic field along the line-of-sight and integrated over the visible stellar hemisphere, $B_{\text{los}}$\footnote{$B_{\text{los}}$ is also referred to as the longitudinal field $B{_\ell}$. We chose not to use this term to avoid confusion with its homonym $B_{\phi}$, the field along the east-west direction or azimuthal field \cite[see e.g.,][]{2016MNRAS.459.1533V}.}, 
by using equation 3.3 from \cite{2012EAS....57..165M}:
\begin{equation}
B_{\text{los}}(G) = -2.14 \times 10^{11} \frac{\int{v \, V(v) \, \text{d}v}}{\lambda_{0} \, g_{\text{eff}} \, c \, \int{[I_{c} - I_{v}] \, \text{d}v}}
\end{equation}
on the LSD profiles of the photospheric absorption lines \cite[see also][]{1979A&A....74....1R, 1997MNRAS.291..658D, 2000MNRAS.313..851W}. In this equation, $v$ is the radial velocity in the rest frame of the star, $V$ refers to Stokes\,V, $\lambda_{0}$ is the wavelength of the line center in nm, $g_{\text{eff}}$ is the effective Landé factor of the line, $I$ refers to Stokes\,I and $I_{c}$ to the unpolarized continuum. 
We find values ranging from -0.19 $\pm$ 0.05 kG to 0.20 $\pm$ 0.03 kG in 2010 and from -0.13 $\pm$ 0.02 kG to 0.08 $\pm$ 0.02 kG in 2012 (see Fig.\,\ref{Blosvsveiling20102012}, and see the list of values in Appendix\,\ref{LSDprof}). 
It should be noted that, since $B_{\text{los}}$ represents a signed average over the visible stellar hemisphere, regions of opposite polarities partly cancel out. 

We applied a phase dispersion minimization \cite[PDM;][]{1978ApJ...224..953S} 
technique on the 2010 $B_{\text{los}}$ values and found a period of 8.20 $\pm$ 0.13 days. Since this is the period of the modulation of the stellar magnetic field, it accurately represents the stellar rotation period. This period is consistent with values found in the literature from photometry (8.4\,days from \citealp{1993AAS..101..485B}, 
and 8.18\,days from \citealp{2010PASP..122..753P} and \citealp{2012AstL...38..783A}). 
Small discrepancies could be explained by differential rotation: different measurements tracking features at various latitudes. 

\begin{figure*}[hbtp]
\centering
\includegraphics[scale=0.57]{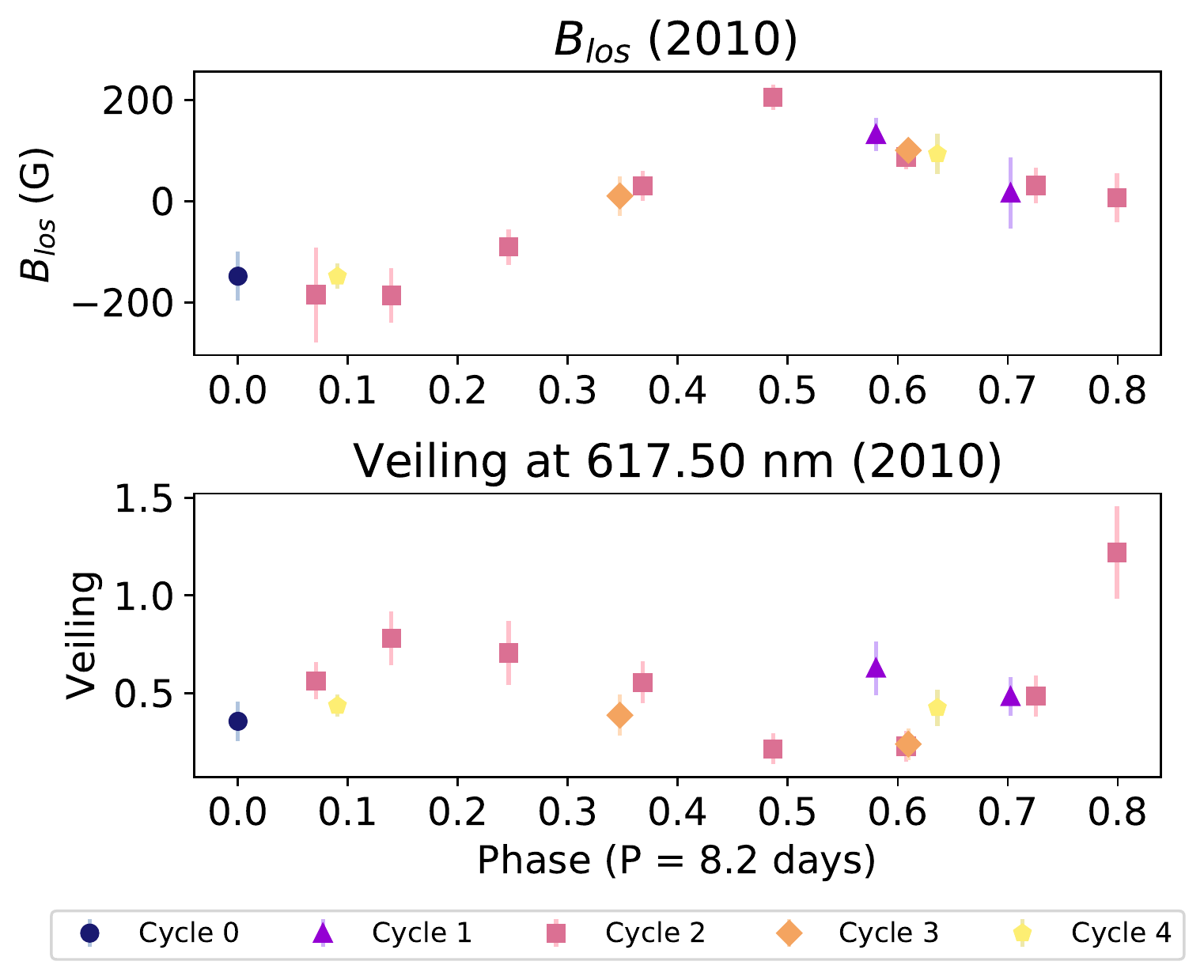}
\includegraphics[scale=0.57]{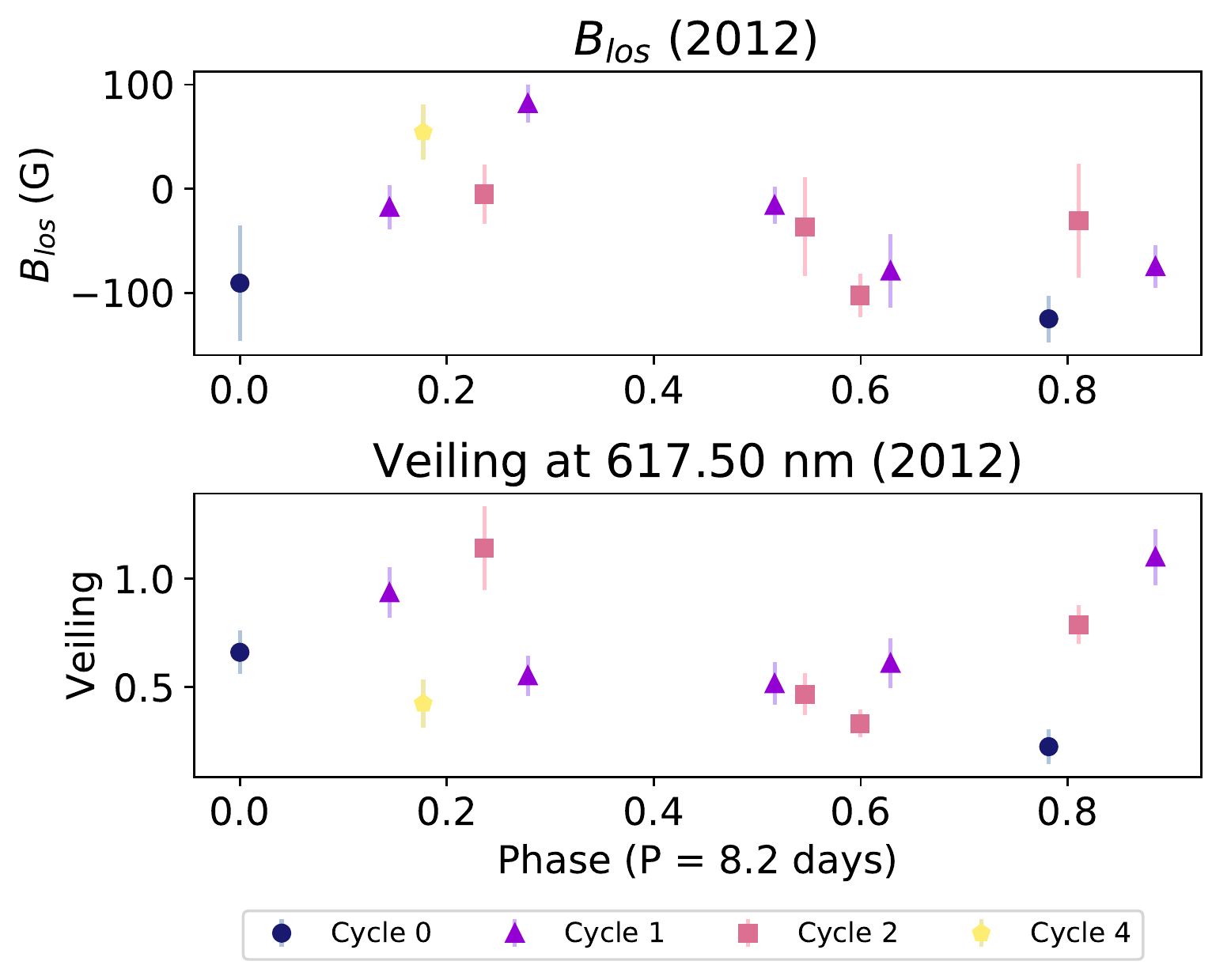}
\caption{Average line-of-sight magnetic field $B_{\text{los}}$ as derived from the photospheric absorption lines and veiling over time, shown folded in phase with the derived 8.2\,day period, for the 2010 (left panel) and 2012 (right panel) datasets. 
Different colors and symbols represent different rotation cycles. 
\label{Blosvsveiling20102012}}
\end{figure*}


\subsection{Emission lines} \label{EL}

After analyzing the LSD profiles of photospheric absorption lines to derive the associated $B_{\text{los}}$ linked to non-accreting regions, we also investigated emission lines associated with accretion shocks, in particular the narrow component 
of the 587.6\,nm He{\sc i} line and of the Ca{\sc ii} infrared triplet (IRT - at 849.8\,nm, 854.2\,nm and 866.2\,nm). 
They can be used to get more information on DK\,Tau's magnetic fields, as they are tracers of the fields present at the footpoints of accretion funnels \cite[see e.g.,][]{2007MNRAS.380.1297D}. 
These lines are known to have multiple components \cite[see e.g.,][]{2001ApJ...551.1037B, 2020MNRAS.497.2142M}, 
often containing a narrow component (NC), originating from the accretion shock, and a much broader component (or components), which likely forms farther out in the accretion columns or in hot winds. For this reason, we decomposed the profiles using a fit of 2 or more Gaussians in order to isolate the NC (examples of these fits can be seen in Appendix\,\ref{DecompPlots}). 
For the Ca{\sc ii} lines, we then averaged the three lines into a single LSD-like profile in order to increase the signal to noise ratio, as it has been done in other studies \cite[see e.g.,][]{2007MNRAS.380.1297D, 2008MNRAS.386.1234D, 2012MNRAS.425.2948D}. 
Since this is a triplet, the shape of all three lines should be the same \cite[see e.g.,][]{2006A&A...456..225A}. In addition, in our data, we see an intensity ratio close to 1:1:1 and the NCs are not contaminated by the nearby Paschen emission lines at  850.2\,nm, 854.5\,nm and 866.5\,nm. 
The NC of the He{\sc i} line is believed to be generated in the post-shock region at the base of the magnetic accretion funnels that connect the surface of the star to its inner disk. The NC of the Ca{\sc ii} IRT is thought to probe the accretion regions and the chromosphere. 

\begin{figure*}[hbtp]
\centering
\includegraphics[scale=0.45]{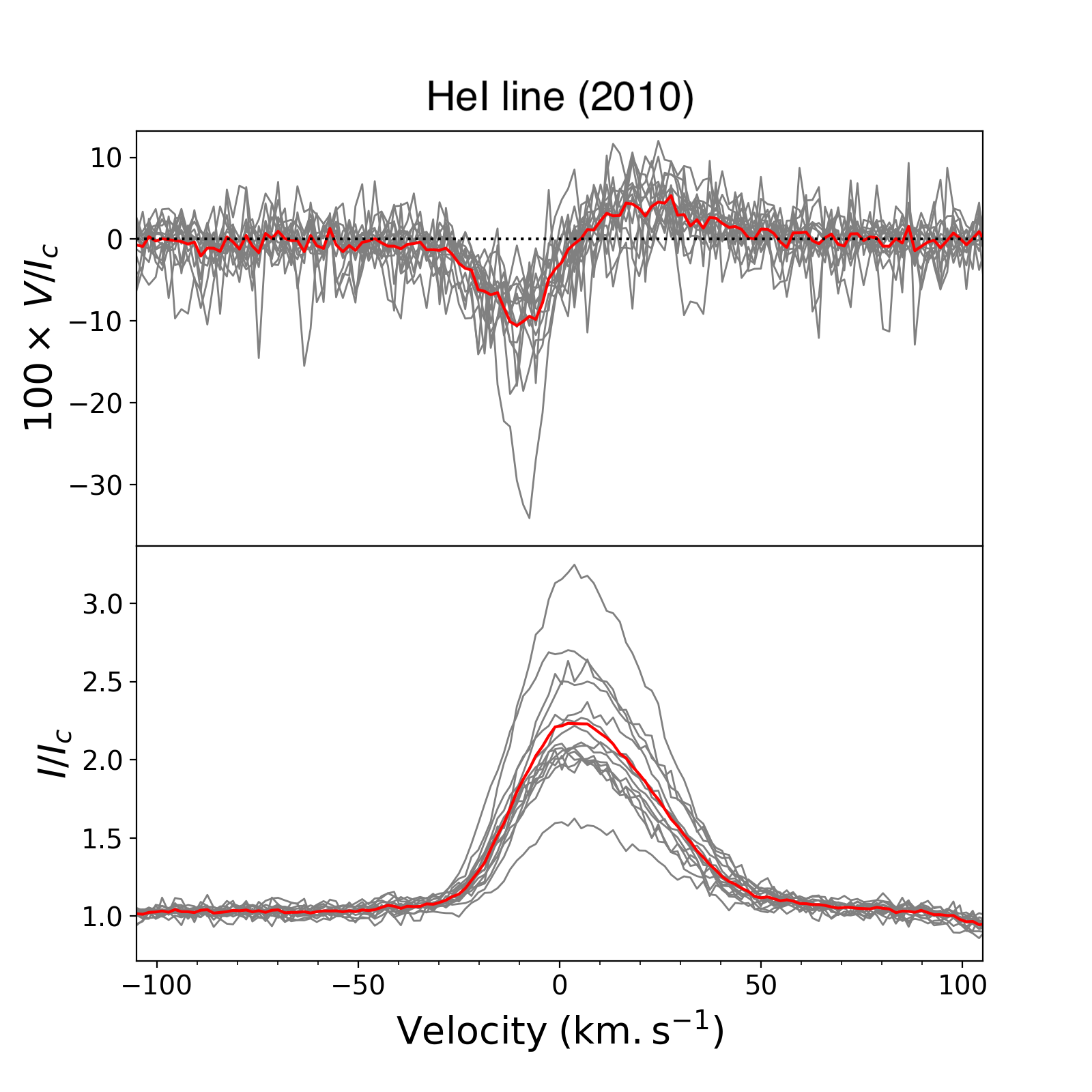}
\includegraphics[scale=0.45]{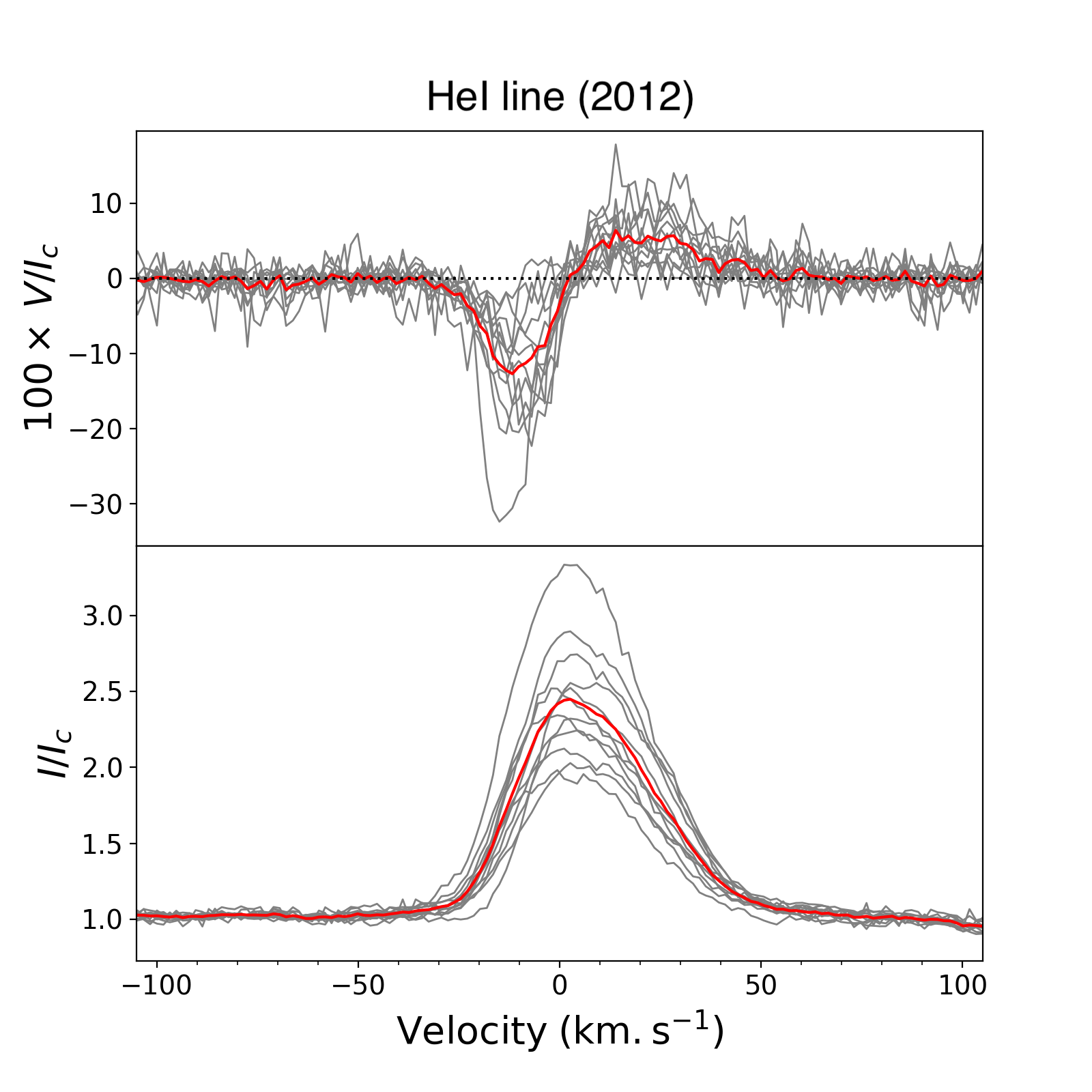}
\includegraphics[scale=0.45]{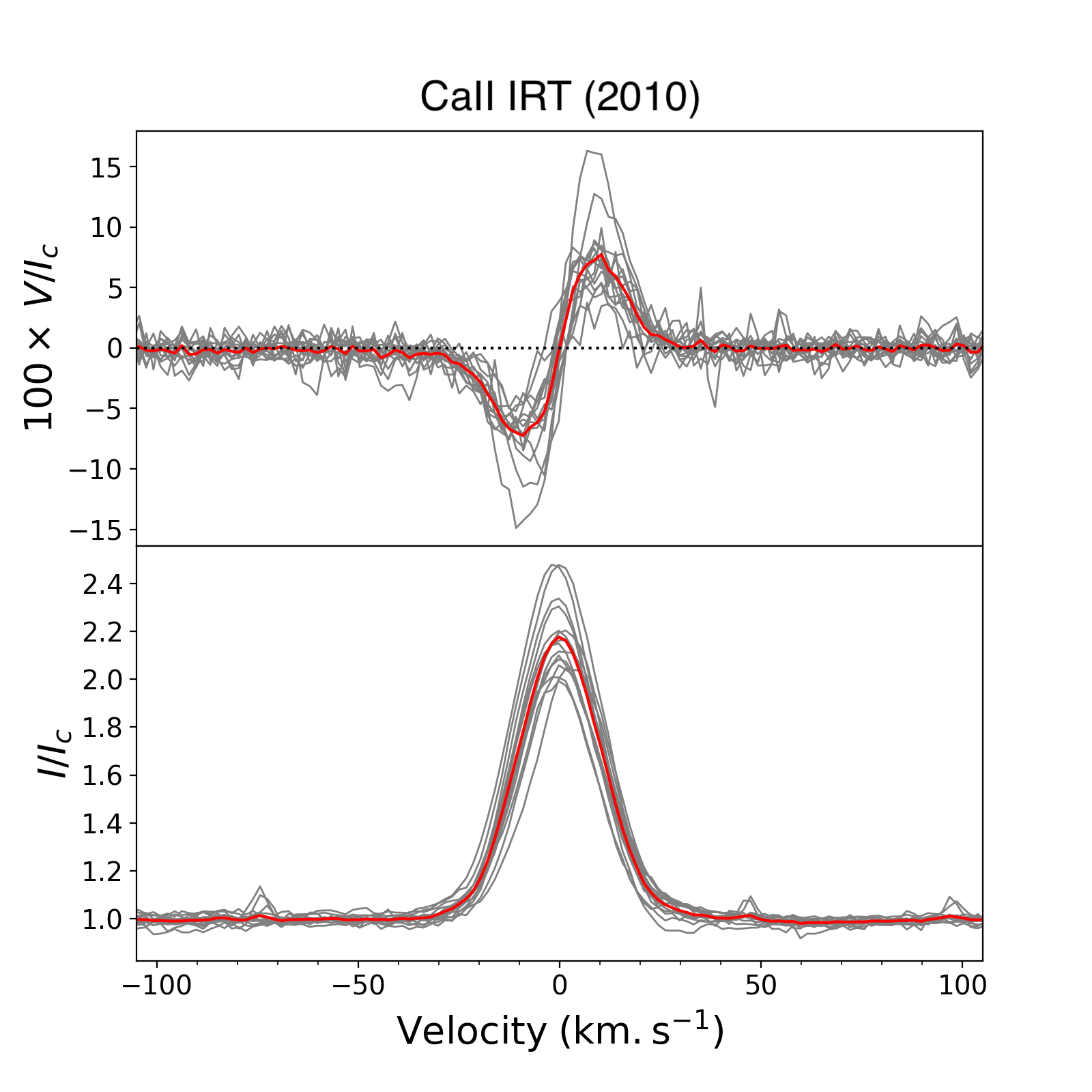}
\includegraphics[scale=0.45]{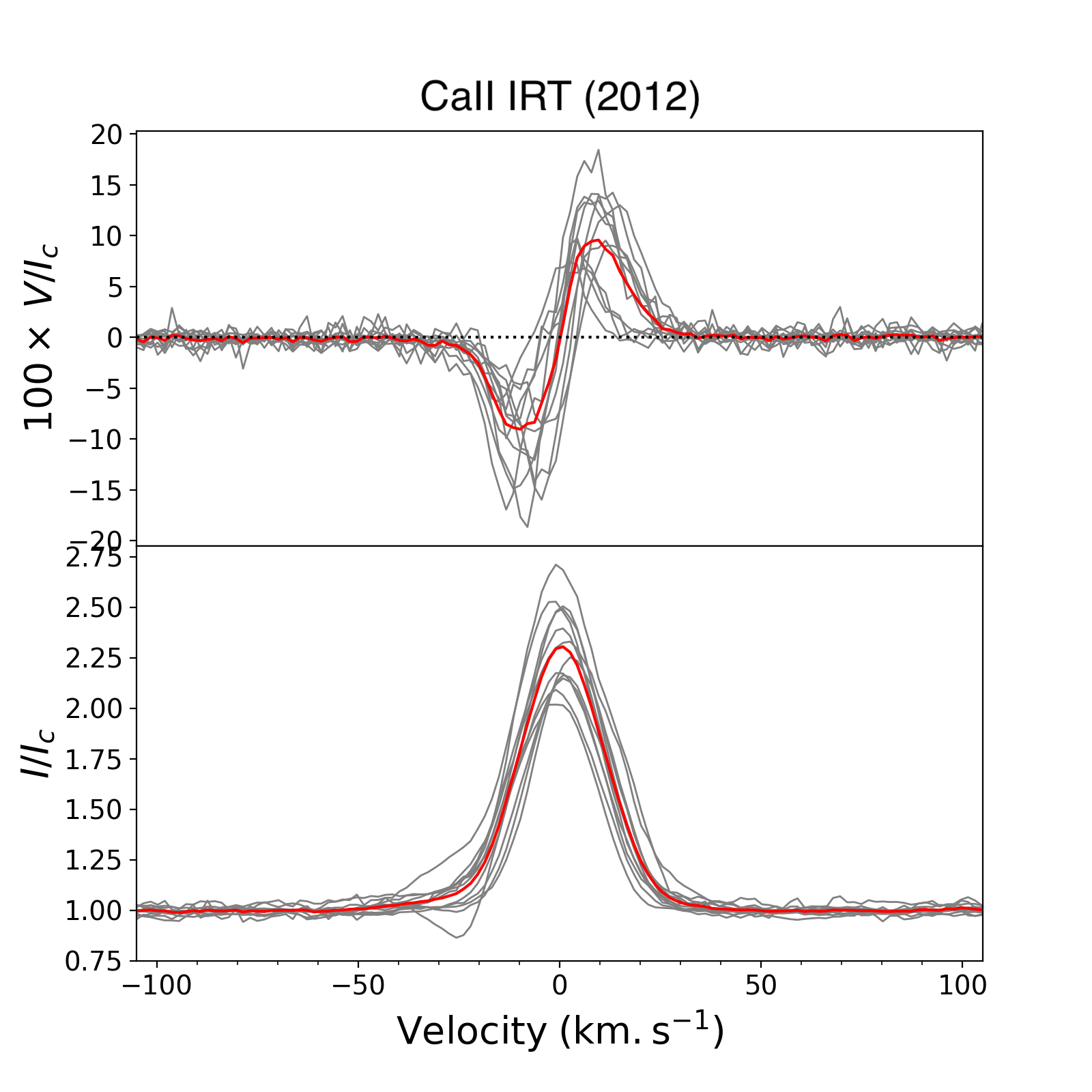}
\caption{Stokes\,V and Stokes\,I profiles (in gray) and average (in red) of the He{\sc i} emission line (top panels) and the Ca{\sc ii} IRT (bottom panels), for the 2010 (left panels) and 2012 (right panels) observations. \label{ELPlots}}
\end{figure*}

The Stokes\,V and Stokes\,I profiles of the He{\sc i} emission line and the Ca{\sc ii} IRT, for both epochs, can be seen in Fig.\,\ref{ELPlots}. The Stokes\,V profiles of the emission lines show similar signatures with phase. This indicates that the accretion spot is mostly likely located at a high enough latitude to be visible with the same polarity at all times. They also indicate that the field is positive (i.e., pointing toward the observer) at the base of the accretion funnels that connect the star to its circumstellar disk. 

We measured $B_{\text{los}}$ in the emission lines using the same method as described in Sect.\,\ref{Blos}. We find values ranging from 0.20 $\pm$ 0.51 kG to 1.77 $\pm$ 0.08 kG for He{\sc i} in 2010, from 0.48 $\pm$ 0.09 kG to 1.99 $\pm$ 0.09 kG for He{\sc i} in 2012, from 0.34 $\pm$ 0.04 kG to 0.95 $\pm$ 0.02 kG for Ca{\sc ii} in 2010, from 0.26 $\pm$ 0.02 kG to 1.41 $\pm$ 0.05 kG for Ca{\sc ii} in 2012.  Figure\,\ref{BlosEL} shows the obtained values folded in phase (and the list of values can be found in Appendix\,\ref{DecompPlots}). The values of $B_{\text{los}}$ are higher in 2012. 
We find that the values measured through the Ca{\sc ii} IRT are lower than the ones measured through the He{\sc i} line. 
It has been hypothesized that the lower intensity of $B_{\text{los}}$ found using the Ca{\sc ii} IRT stems from the dilution of the emission from the accretion shock by chromospheric emission \cite[see e.g.,][]{2019MNRAS.483L...1D}, which results in the Stokes\,V/I profile being shallower for the Ca{\sc ii} IRT than for the He{\sc i} line. 

\begin{figure*}[hbtp]
\centering
\includegraphics[scale=0.56]{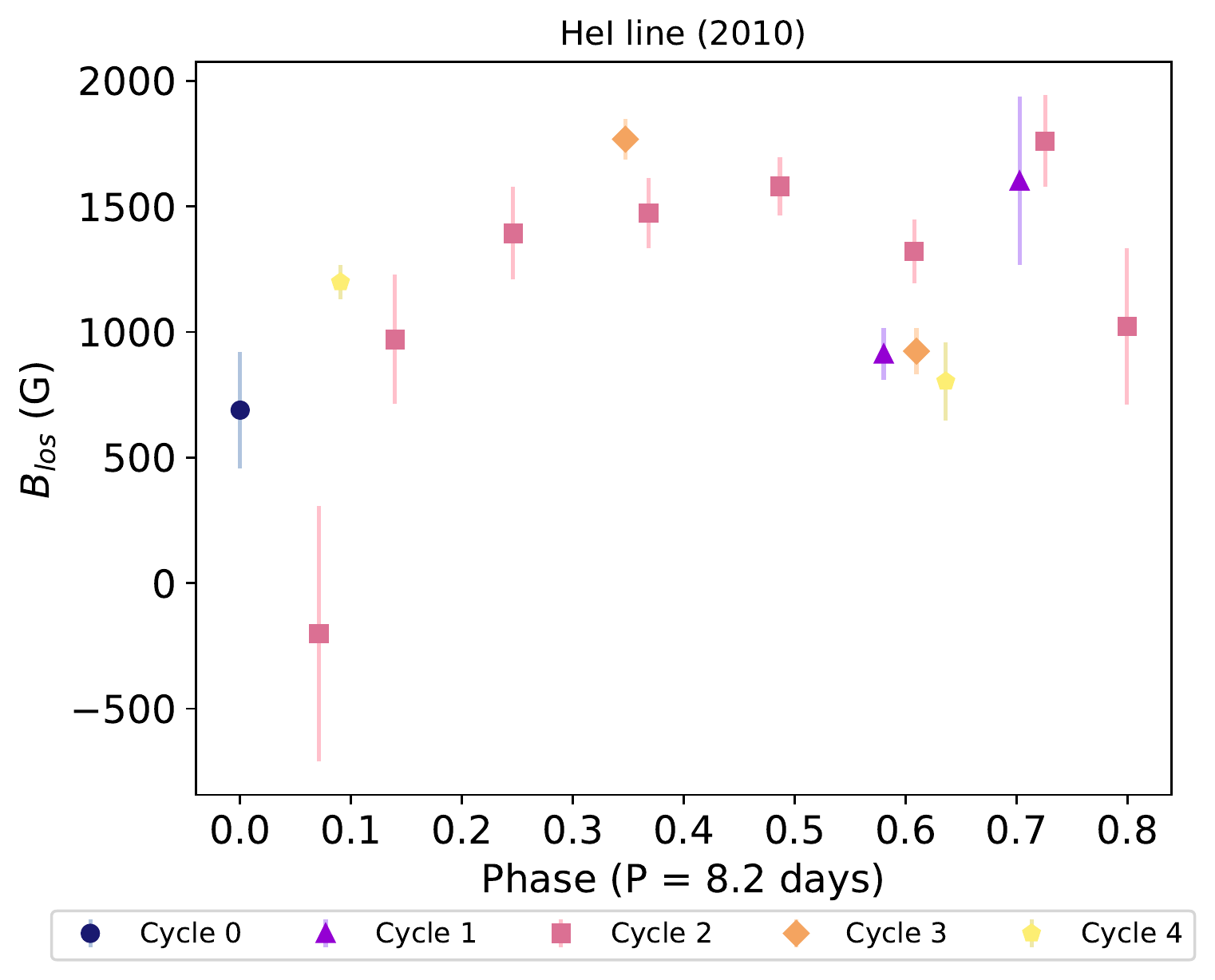}
\includegraphics[scale=0.56]{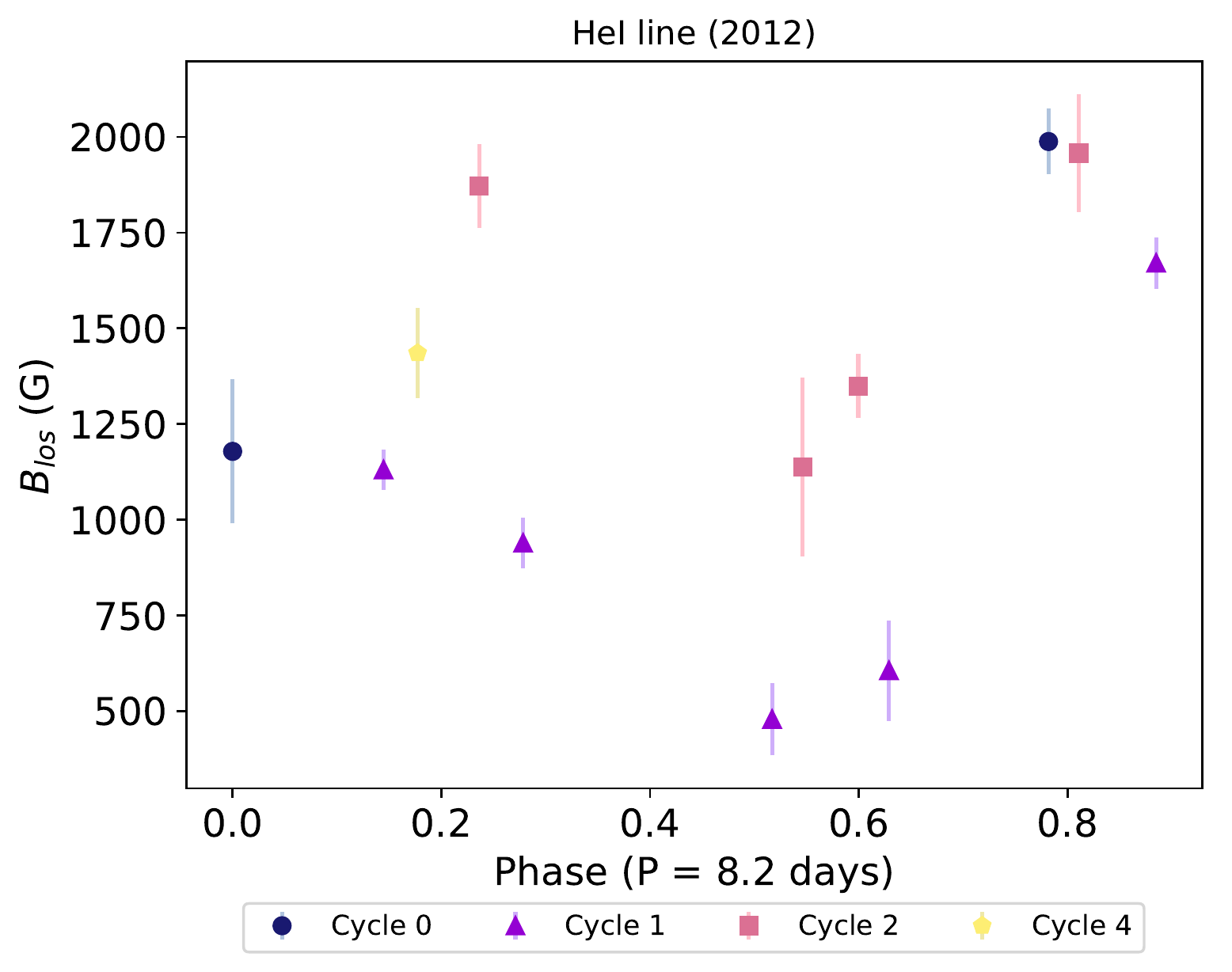}
\includegraphics[scale=0.56]{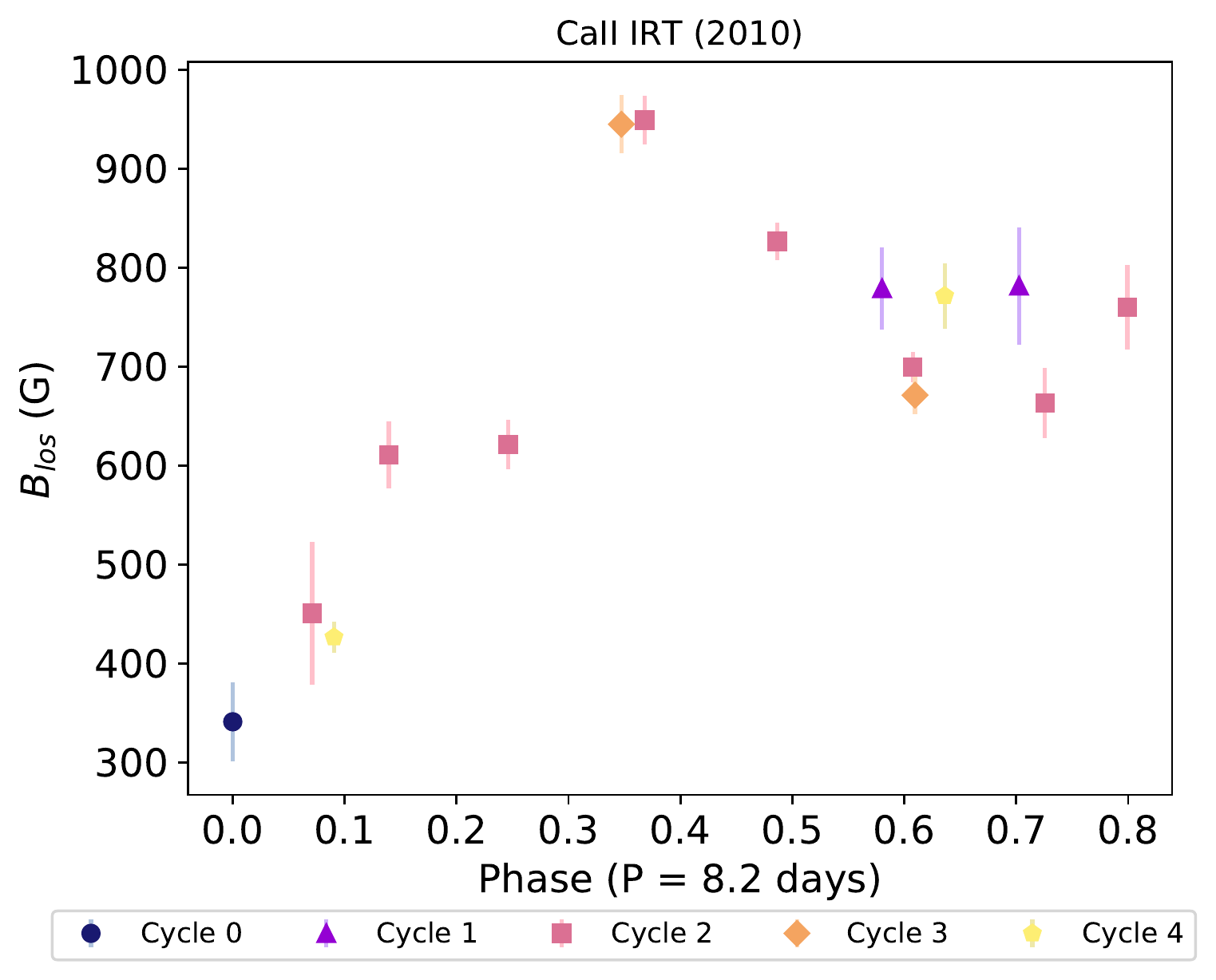}
\includegraphics[scale=0.56]{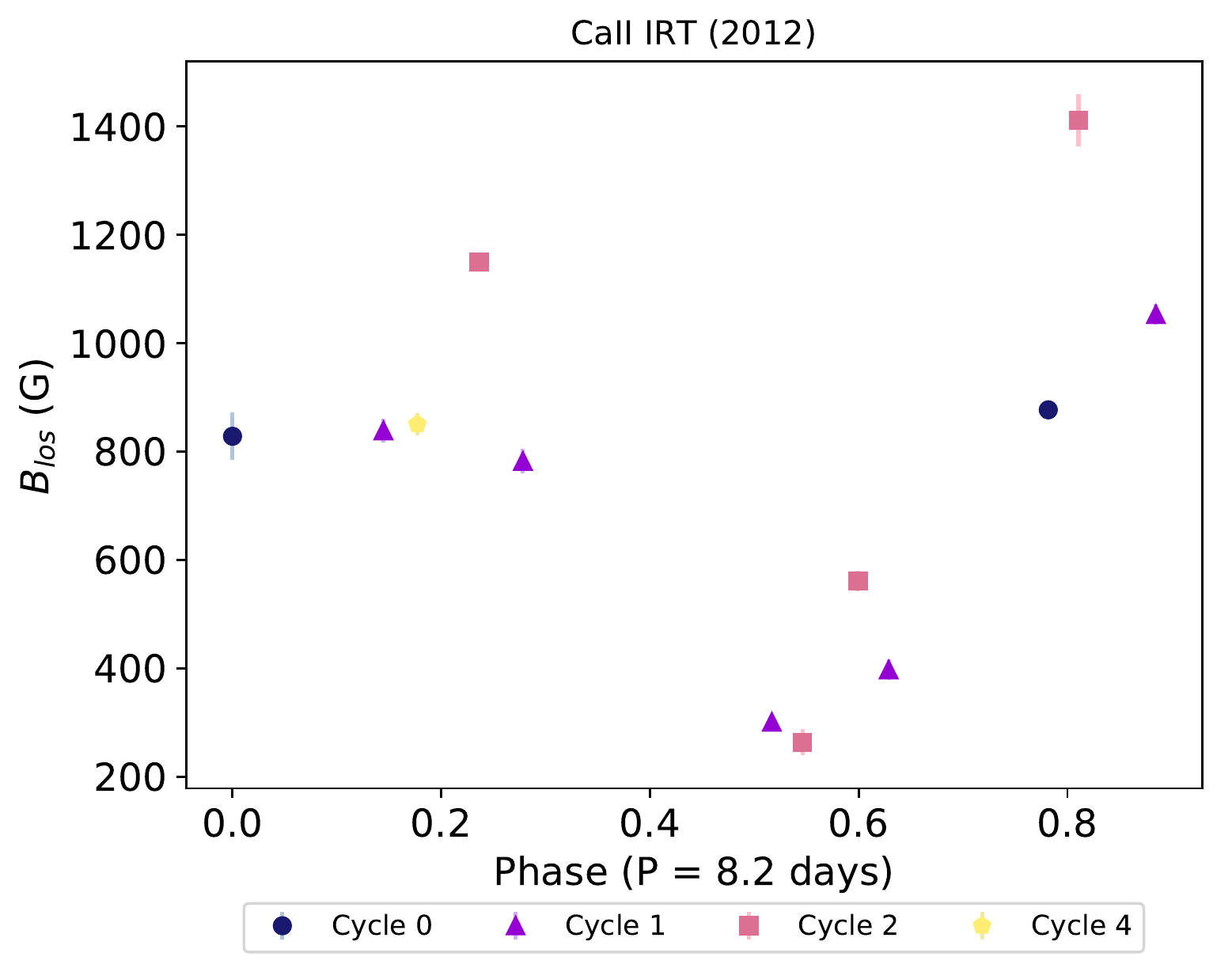}
\caption{Average line-of-sight magnetic field $B_{\text{los}}$ over time, shown folded in phase with an 8.2\,day period, for the He{\sc i} emission line (top panels) and the Ca{\sc ii} IRT (bottom panels), for the 2010 (left panel) and 2012 (right panel) datasets. Different colors and symbols represent different rotation cycles. \label{BlosEL}}
\end{figure*}

We find extreme values of $B_{\text{los}}$ in the emission lines that are one order of magnitude larger than the extreme values of $B_{\text{los}}$ derived from the LSD profiles of the photospheric absorption lines (see Fig.\,\ref{Blosvsveiling20102012}), showing strong fields in the accretion shocks. This is consistent with the current understanding of accretion shocks as compact regions with some of the strongest magnetic field concentrating in dark polar regions at the surface of the star and magnetic field lines reaching to the circumstellar disk. 
For both epochs, we only see the positive pole and never the negative one. 

The plots of the $B_{\text{los}}$ in the emission lines in 2012 do not fold well in phase. We believe this may stem from the location of the accretion shocks being more dynamic and/or the accretion being more complex, with more than one accretion shock. This is also observed in the variability of veiling in 2012 (see Fig.\,\ref{Blosvsveiling20102012}), which does not fold well with the rotation phase, indicating that there is considerable intrinsic variability in the mass accretion rate. 

We derived the equivalent width (EW) of the He{\sc i} emission line. Figure\,\ref{EWHeI} shows the obtained values folded in phase (and the list of values can be found in Appendix\,\ref{DecompPlots}). When the EW is larger, we are seeing more of the accretion shock in our line-of-sight. This is also when we measure stronger magnetic fields using emission lines tracing the accretion shock (see Fig.\,\ref{BlosEL}), as expected following the paradigm of magnetospheric accretion. 

\begin{figure*}[hbtp]
\centering
\includegraphics[scale=0.56]{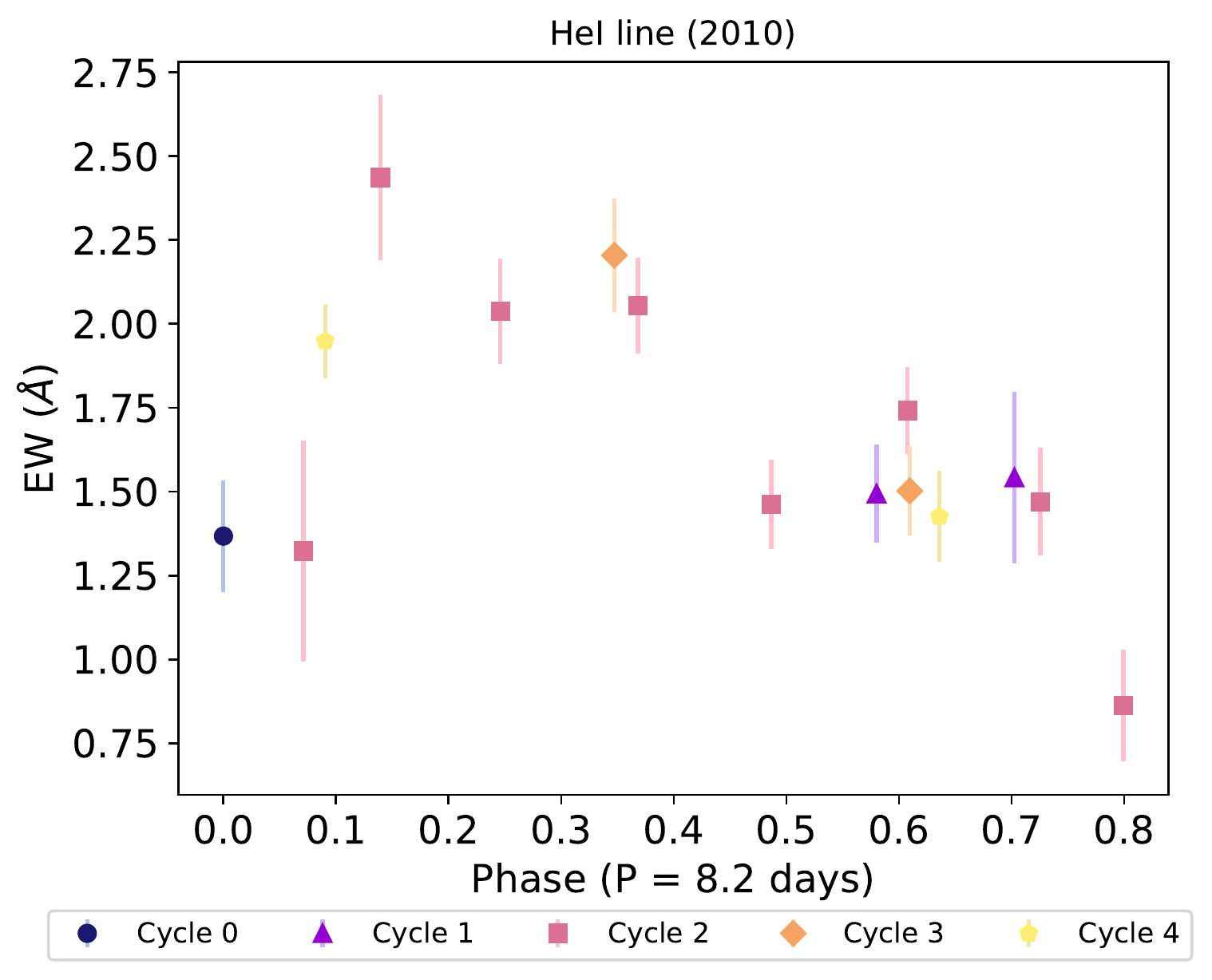}
\includegraphics[scale=0.56]{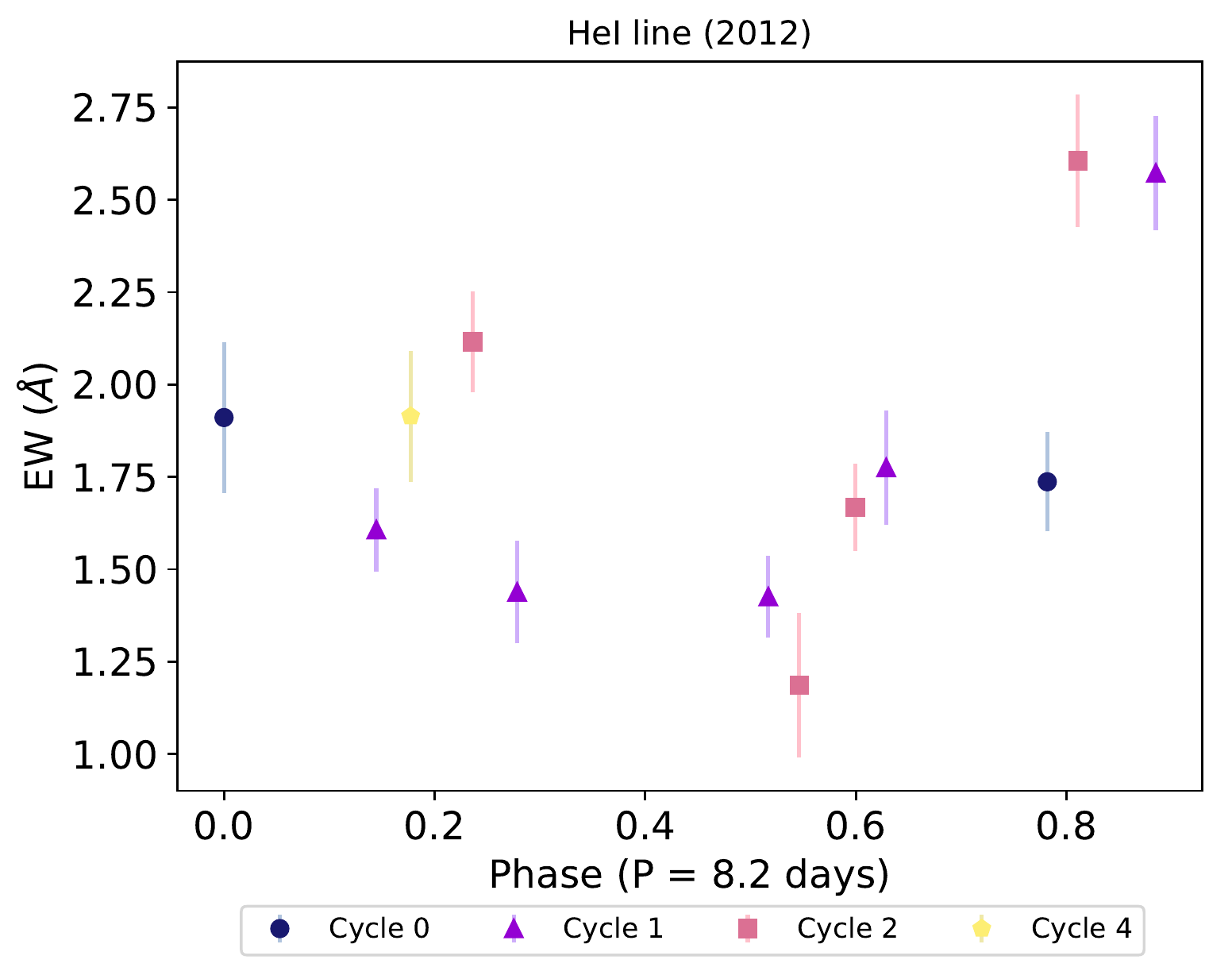}
\caption{Equivalent width (in $\mathring{A}$) of the He{\sc i} emission line over time, shown folded in phase with an 8.2\,day period, for the 2010 (left panel) and 2012 (right panel) datasets. Different colors and symbols represent different rotation cycles. \label{EWHeI}}
\end{figure*}


\subsection{Magnetic Obliquity} \label{Obliquity}

We used the third equation of \cite{1967ApJ...150..547P}, which assumes a pure dipole, a simplification of the magnetic field present in the accretion shocks, to calculate an estimate of the magnetic obliquity (i.e., the angle between the stellar rotation axis and the magnetic field axis) derived from the emission lines. 
We used the extreme values found for $B_{\text{los}}$ in the emission lines and an inclination $i$ of 58\si{\degree}.
For the He{\sc i} emission line, we find a magnetic obliquity\footnote{This is not the magnetic obliquity of the entirety of DK\,Tau's magnetic field. It is based solely on the average line-of-sight magnetic field present in the accretion shocks and probed through emission lines. Furthermore, the calculation uses the extreme values for $B_{\text{los}}$ and does not account for variability between nights.} of 26\si{\degree} for the 2010 epoch, and of 21\si{\degree} for the 2012 epoch. For the Ca{\sc ii} IRT, we find a magnetic obliquity of 16\si{\degree} for the 2010 epoch, and of 23\si{\degree} for the 2012 epoch. These estimates are consistent with the Stokes\,V signatures of the emission lines. 
They are also consistent with the magnetic obliquity of 18\si{\degree} (+8)(-7) in 2011 derived by \cite{2020MNRAS.497.2142M}, using the radial velocity variability of the He{\sc i} emission line and assuming one accretion spot. 
We therefore have an agreement between the values derived from the magnetic field that drives the accretion and the value derived from a result of accretion. 

The estimates of the magnetic obliquity are consistent with only seeing the positive pole in the plots of the $B_{\text{los}}$ in the emission lines, confirming that DK\,Tau experiences nearly poleward accretion, with a positive field at the base of the accretion funnels. 

The He{\sc i} emission lines show a radial velocity variability with a small amplitude, which is another indication that the accretion spot is most likely close to the pole. As the star rotates, if the accretion spot were located at the equator, the velocity variation would be large as the spot gets red and blueshifted. 


\subsection{Truncation \& co-rotation radii} \label{TruncCorot}

In order to calculate the truncation radius, we need to know the star's mass accretion rate. For this, we measured the equivalent width (EW) of several emission lines (i.e., H$\alpha$, H$\beta$, H$\gamma$, the He{\sc i} lines at 447.1\,nm, 667.8\,nm and 706.5\,nm, as well as the Ca{\sc ii} IRT at 849.8\,nm, 854.2\,nm and 866.2\,nm). 
Since ESPaDOnS and NARVAL's spectra are not flux calibrated, 
we created a template for DK\,Tau based on SO879, a weak-lined T\,Tauri star with a K7 spectral type \cite[described in][]{2013A&A...558A.141S}. 
The template was corrected for extinction, then scaled to have the same luminosity (i.e., 1.65\,$L_{\odot}$) and be at the same distance (i.e., 132.6\,pc) as DK\,Tau. After correcting the EW for veiling, we used this template to flux calibrate them through the following formula: 
 \begin{equation}
F_{\text{line}} = EW_{\text{line}} \cdot F_{\text{cont}}
\label{FormulaFluxCal}
\end{equation} 
where $F_{\text{line}}$ is the flux of the line, $EW_{\text{line}}$ is the veiling corrected EW of the line and $F_{\text{cont}}$ is the flux of the continuum of the template at the wavelength of the line in question. 
Then we obtained the luminosity in each line. 
Next, we used the relations in Table\,B.1. from \cite{2017A&A...600A..20A} 
to calculate the accretion luminosity from each line 
and averaged these values for each night. 
We then took the average over all nights in each epoch as the accretion luminosity, and the standard deviation of the spread in the values found from different nights as the error bars. 
We find $L_{\text{acc}}$ = 0.26 $\pm$ 0.18 $L_\odot$ in 2010 and $L_{\text{acc}}$ = 0.49 $\pm$ 0.42 $L_\odot$ in 2012. 
These values are similar to the ones found e.g., by \cite{2011ApJ...730...73F} 
(i.e., 0.17\,$L_\odot$) or by \cite{2018ApJ...868...28F} (i.e., 0.16\,$L_\odot$). 

We then converted the accretion luminosity into mass accretion rate using Eq.\,8 from \cite{1998ApJ...492..323G} with the values of $R_\star$ and $M_\star$ from Table\,\ref{TableProp} and $R_{\text{in}}$ = 5 $R_\star$ (as is typically used). 
We find log\,($\dot{M}_{\text{acc}}$[$M_\odot$\,yr$^{-1}$]) = -7.43 in 2010, and log\,($\dot{M}_{\text{acc}}$[$M_\odot$\,yr$^{-1}$]) = -7.15 in 2012. 
These values are consistent with the one of -7.42 quoted by \cite{1998ApJ...492..323G}. 
Finally, we used Eq.\,6 from \cite{2008A&A...478..155B} to estimate the truncation radius. 
This equation assumes an axisymmetric dipole, which is a simplification of DK Tau's magnetic topology. 
It also uses the dipolar field calculated at the equator as $B_\star$. Considering the equatorial value is half of the value at the pole, 
we estimated the latter \cite[see][]{1967ApJ...150..547P} using the values of $B_{\text{los}}$ in the emission lines \footnote{We used the magnetic field derived from the accretion-powered emission lines, considering it will dominate over the magnetic field derived from the photospheric absorption lines at the distance of the truncation radius. }. 
This is an approximation, as part of the $B_{\text{los}}$ could come from higher order multipoles, in particular from the octupole, rather than the dipole. 
We find $r_{\text{trunc}}$ $\sim$ (5.2 $\pm$ 1.1) $R_\star$ for the He{\sc i} emission line in 2010, 
$r_{\text{trunc}}$ $\sim$ (3.9 $\pm$ 0.8) $R_\star$ for the Ca{\sc ii} IRT\footnote{We find lower values for the truncation radius estimates when using the Ca{\sc ii} IRT. This stems from the lower values found for $B_{\text{los}}$ in those lines (see Sect.\,\ref{EL}).} in 2010, 
$r_{\text{trunc}}$ $\sim$ (4.7 $\pm$ 1.2) $R_\star$ for the He{\sc i} emission line in 2012, 
and $r_{\text{trunc}}$ $\sim$ (3.9 $\pm$ 1.0) $R_\star$ for the Ca{\sc ii} IRT in 2012. 
We calculated the co-rotation radius as well, using Kepler's third law: $r_{\text{co-rot}}$ = 6.1\,$R_\star$.
We find that the truncation radius values are consistent with the co-rotation radius within the error bars. 
This implies that DK\,Tau is unlikely to be in the propeller regime, an unstable accretion regime, as the truncation radius is not farther than the co-rotation radius \cite[][]{2018NewA...62...94R}.
For the 2010 epoch, DK\,Tau may be in the stable accretion accretion regime, since the $B_{\text{los}}$ and accretion tracers seem fairly periodic. The truncation radius being slightly smaller than the co-rotation radius is consistent with this as well \cite[][]{2016MNRAS.459.2354B}. 


\section{Discussion} \label{Disc}

\subsection{Inconsistencies regarding the inclination} \label{ibeta}

Naively one might assume that the inclination angle $i$ of the stellar rotation axis with respect to the line-of-sight is 21\si{\degree} based on the inclination of the outer gaseous disk axis \cite[][]{2022arXiv220103588R}. 
DK\,Tau's lightcurve however classifies the star as a dipper \citep{2021A&A...651A..44R}. 
The traditional explanation invokes circumstellar material passing in front of the star and occulting it. If the disk is seen close to edge on, matter lifted above the disk plane could cause these occultations. However, DK\,Tau's outer disk is seen nearly pole on \citep{2022arXiv220103588R}, which is inconsistent with this scenario, unless the stellar rotation axis is at a very different angle than that of the outer disk axis.

Furthermore, based on the star's rotational properties (see Table\,\ref{TableProp}) and using the following relation 
\begin{equation}
v \sin i = \frac{2 \pi R_\star}{P} \sin i
\label{EquationR}
\end{equation} 
we derive a much higher inclination of 58\si{\degree} (+18)(-11). 
Therefore, if DK\,Tau is in fact seen nearly pole on, then $P$, $v \sin i$ and $R_\star$ are not consistent with each other. As the stellar radius is the most uncertain of these parameters\footnote{This is because it depends on evolutionary models as well as an accurate determination of the effective temperature and stellar luminosity. Both can be subject to fairly large uncertainties, particularly the luminosity for a star with a dipper light curve which likely suffers from variable extinction.}, it is possible that it may have been underestimated. However, if we consider that $P$, $v \sin i$ and $i$ are accurately determined, then we would need $R_\star$ = (6 $\pm$1) $R_\odot$ for this formula to agree, which is unrealistically large for a TTs. 

Another possibility is that the period or $v \sin i$ may be inaccurate. Regarding $v \sin i$, the value we derive agrees within error bars with the one measured by \cite{2020MNRAS.497.2142M}, despite using two different assessment methods. It is therefore a value that can be trusted. This leaves the stellar rotation period.

In the literature, the stellar rotation period of DK\,Tau has been measured using photometry with values ranging from 8.18\,days  (\citealp{2010PASP..122..753P,2012AstL...38..783A}) to 8.4\,days  (\citealp{1993AAS..101..485B}). 
However, since the photometry is dominated by flux dips that might be due to extinction events \citep{2021A&A...651A..44R}, it is possible that these dips are caused by circumstellar material that is not located at the co-rotation radius. In that case, the measured period would not be the same as the stellar rotation period. 

In Sect. \ref{Blos} we derived a period from the rotational modulation of the line-of-sight magnetic field $B_{\text{los}}$, which should accurately represent the stellar rotation period. 
In the context of exoplanet search programs, $B_{\text{los}}$ is indeed often considered as the most reliable indicator of stellar rotation period \cite[see e.g.,][]{2016MNRAS.461.1465H}. 
Additionally, the value we find of 8.20 $\pm$ 0.13 days is consistent with those found in the literature from photometry. 
We therefore find that the value for the period can be trusted. 

Moreover, this rotation period can be seen in a number of datasets at our disposal. For example, we computed bidimensional periodograms of the intensity of the He{\sc i} (at 587.6\,nm) emission line, and the Stokes\,I and Stokes\,V LSD profiles of the photospheric absorption lines (see Appendix\,\ref{2DPer}).  
The He{\sc i} line comes from the accretion shock and should therefore vary with the stellar rotation period. 
In 2010 we find a period around 8 days for the entire red-shifted part of the He{\sc i} line (from 0 to 50\,km\,s$^{-1}$), however this period is very uncertain. The Stokes\,V profile also shows a period near 8.5 days between $\sim$-20 and -7\,km\,s$^{-1}$ and between $\sim$0 and 8\,km\,s$^{-1}$, but again the uncertainty is large. The Stokes\,I profile does not show a clear period. 
In 2012 there is no clear period found from the He{\sc i} line, likely because accretion is more intrinsically variable in this epoch than 2 years prior. Again no clear period is observed from the Stokes\,I profile, but the Stokes\,V profile shows a possible periodicity at around 8 days (albeit with a large uncertainty, same as in 2010).

Furthermore, the variation of the 2010 veiling as a function of time is also consistent with an 8.2\,day period (see Fig.\,\ref{Blosvsveiling20102012}). The variation of the 2012 veiling as a function of time, however, does not seem to follow any trend with the period. This is consistent with the intensity of the He{\sc I} line not showing a clear correlation with period in this epoch, since both are tracing accretion. This is another indication that there must be some intrinsic variability in the mass accretion rate in 2012, which masks any rotational modulation of the accretion spot(s).

\begin{figure}[hbtp]
\begin{center}
\includegraphics[scale=0.34]{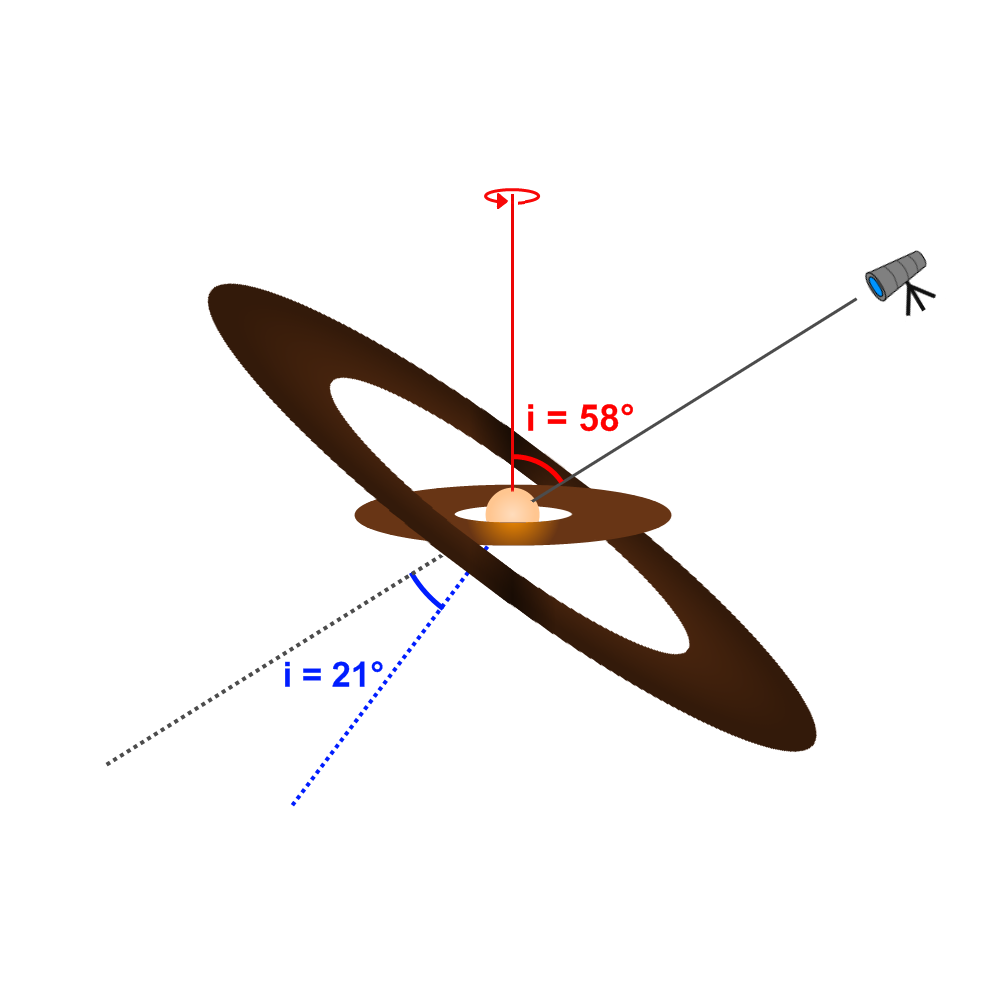}
\caption{Sketch (not to scale) showing DK\,Tau in the center, surrounded first by its inner disk, then by its outer disk which is considerably misaligned. The rotation axis at 58\si{\degree} is in red, the outer disk axis at 21\si{\degree} is in blue, and the line-of-sight axis is in gray (by C. Delvaux).\label{Artists_Vision}}
\end{center}
\end{figure}

We looked at the possibility of the rotation period or $v \sin i$ being inaccurate and found evidence to the contrary. Because their values appear to be accurate, we deduce that it is the value for the inclination that is problematic. 
We conclude that the inclination measured for the outer circumstellar disk axis must not represent the inclination of the rotation axis of the star. This suggests that there is a considerable misalignment between the rotation axis of DK\,Tau and its outer disk. 
When we calculate DK\,Tau's inclination based on its rotational properties using Eq.\,\ref{EquationR}, we find $i$ = 58\si{\degree} (+18)(-11). This value is based on $v \sin i$ = (13.0  $\pm$ 1.3) km\,s$^{-1}$, $P$ = (8.2 $\pm$ 0.2) days and $R_\star$ = (2.48 $\pm$ 0.25) $R_\odot$. It follows that the outer disk axis of DK\,Tau is likely misaligned by 37\si{\degree} with its rotation axis (see Fig.\,\ref{Artists_Vision}). 

Misalignments between the inner and outer circumstellar disk axes of T Tauri stars are starting to be observed, when combining near infrared interferometric VLTI/GRAVITY data and millimeter interferometric ALMA data \cite[see e.g.,][]{2020MNRAS.492..572A, 2020A&A...636A.108B}, 
or with shadows observed with VLT/SPHERE \cite[see e.g.,][]{2018A&A...619A.171B, 2020A&A...633A..37S}, 
or with VLTI/GRAVITY \cite[see e.g.,][]{2022A&A...658A.183B}.
We find a misalignment between the outer disk axis and the rotation axis. What of the inner disk axis? 
In young dippers like DK\,Tau, the material that causes the dips is believed to be located in the inner accretion disk \cite[][]{2007prpl.conf..479B, 2015A&A...577A..11M}, 
which need to be observed at high inclinations in order for material to cross our line-of-sight to produce the dips. Therefore an inclination of 21\si{\degree} of the inner disk of DK Tau is difficult to reconcile with its dipper light curve. 
In addition, the inner disk axis is normally expected to be aligned with the stellar rotation axis. 
We thus find it very likely that the inner disk axis of DK\,Tau has the same inclination of $i=58\si{\degree}$ as was calculated for its rotation axis. 
This inclination is sufficiently high to support the dipper behavior and there are other cases of dippers with similar inclinations \cite[see e.g.,][]{2015A&A...577A..11M, 2021A&A...651A..44R}. 
DK\,Tau is one more example of a TTS with a misalignment between its inner and outer circumstellar disk axes. 

It is also a wide binary system, and the misalignment could stem from the binary formation mechanism: turbulent fragmentation might generate disk axis that are more randomly oriented. 
It is however interesting to note as well that the outer disk axes of both components of the binary are misaligned by 43\si{\degree} \cite[see][]{2022arXiv220103588R}, 
which is close to the value (of 37\si{\degree}) of the misalignment between the inner and outer disk axes of DK\,Tau~A. 
This could suggest a quasi-alignment of the inner disk axis of DK\,Tau~A with the outer disk axis of DK\,Tau~B, assuming that they are not only aligned compared to our line-of-sight, but that the orientation of their nodes are aligned as well, which is unknown. 


\subsection{Magnetic field in the accretion-powered emission lines}

The $B_{\text{los}}$ derived from the photospheric absorption lines (see Sect. \ref{Blos}), gives a partial view of DK\,Tau's magnetic fields that exclude the accreting regions, as photospheric absorption lines and accretion-powered emission lines form in different regions of the stellar surface. 
It is the field present in these accreting regions that is understood to best probe the global stellar magnetic field that reaches to the circumstellar disk. 
The magnetic obliquity derived from the accretion-powered emission lines is therefore likely to be close to the actual value. We find that the low magnetic obliquity that we derive (see Sect.\,\ref{Obliquity}), the positive polarity of the $B_{\text{los}}$ in the emission lines, as well as their Stokes\,V signatures and the range of radial velocity of the He{\sc i} line are consistent with the presence of an accretion spot always visible and close to the pole. This is where the accretion funnels connecting DK\,Tau to its disk would be anchored. 
This is similar to what has been found for several other cTTs \cite[see e.g.,][]{2014MNRAS.437.3202J, 2020MNRAS.497.2142M}. 

For the 2010 epoch, we find that the magnetic field in the Ca{\sc ii} IRT (and in the He{\sc i} line - see Fig.\,\ref{BlosEL}) is at a maximum close to the same phase (i.e., around phase 0.3) as the maximum in the veiling (see Fig.\,\ref{Blosvsveiling20102012}), which is when the accretion shock is in our line-of-sight. 
Around phase 0.5, we see a small redshifted absorption in the H$\alpha$ line (see Appendix\,\ref{Halpha}), indicating that the accretion column is in our line-of-sight, which is directly after the maximum of the magnetic field in the emission lines and the increase in veiling, therefore probably directly after the accretion shock was in our line-of-sight. 
Because this small redshifted absorption is not perfectly simultaneous with the increase in veiling and in the emission lines, it might be an indication of differential rotation. 

    
\section{Conclusions} \label{Concl}

In this paper, we have studied DK\,Tau, a low-mass classical T\,Tauri star (cTTs) with significant veiling (defined as the ratio between the accretion shock flux and the photospheric flux), using dual-epoch spectropolarimetric observations (collected in 2010 and 2012). We derive an effective temperature 
$T_{\textrm{eff}}$ of 4\,150 $\pm$ 110 K and a line-of-sight-projected equatorial rotational velocity $v \sin i$ of 13.0 $\pm$ 1.3 km\,s$^{-1}$, in agreement with the literature. We find peak values of veiling in the optical ($\sim$550\,nm) ranging from 0.2 to 1.8 in 2010, and from 0.2 to 1.3 in 2012. 

We derive the line-of-sight magnetic field integrated over the visible hemisphere $B_{\text{los}}$ from the photospheric absorption lines (linked to non-accreting regions). We find values ranging from -0.19 $\pm$ 0.05 kG to 0.20 $\pm$ 0.03 kG in 2010 and from -0.13 $\pm$ 0.02 kG to 0.08 $\pm$ 0.02 kG in 2012.

We recover a rotation period of 8.2\,days using the values of $B_{\text{los}}$ for the 2010 dataset. We confirmed the period by analyzing the intensity of the He{\sc i} line in 2010, the intensity of the Stokes\,V profiles in both 2010 and 2012,
and the variation of veiling as a function of time in 2010. They are all consistent with an 8.2\,day period. 
This also agrees with the values of period given in the literature from photometry.

We find several inconsistencies related to the inclination of the stellar rotation axis with respect to the line-of-sight $i$.
The measurement of the inclination of the outer circumstellar disk axis gives a value of 21\si{\degree} \cite[][]{2022arXiv220103588R}. 
DK\,Tau's lightcurve, however, classifies it as a dipper \citep{2021A&A...651A..44R}, for which the simplest explanation involves a star seen close to edge on. 
Furthermore, using Eq.\,\ref{EquationR}, we find that the inclination of 21\si{\degree}, the period, $v \sin i$ and the radius are not consistent with each other. 
When using the values of period, $v \sin i$ and stellar radius that we derive to estimate the inclination of the stellar rotation axis, we find a value of $i$ = 58\si{\degree} (+18)(-11). 
We thus find a substantial misalignment between DK Tau’s rotation axis (at 58\si{\degree}) and its outer disk axis (at 21\si{\degree}) to be likely. 

To complement the partial picture of the $B_{\text{los}}$ derived from the photospheric absorption lines, we analyzed  emission lines that are tracers of the magnetic fields present in the accretion shocks. We examined the narrow component of the 587.67\,nm He{\sc i} emission line and of the Ca{\sc ii} infrared triplet (IRT - at at 849.8\,nm, 854.2\,nm and 866.2\,nm). We found that their Stokes\,V profiles show similar signatures with phase, indicating that DK Tau experiences poleward accretion, with a positive field at the base of the accretion funnels connecting the star to its circumstellar disk. 

We measured $B_{\text{los}}$ within the accretion shocks. We find values ranging from 0.92 $\pm$ 0.09 kG to 1.77 $\pm$ 0.08 kG for He{\sc i} in 2010, from 0.48 $\pm$ 0.09 kG to 1.99 $\pm$ 0.09 kG for He{\sc i} in 2012, from 0.42 $\pm$ 0.02 kG to 0.95 $\pm$ 0.02 kG for Ca{\sc ii} in 2010, from 0.30 $\pm$ 0.01 kG to 1.15 $\pm$ 0.02 kG for Ca{\sc ii} in 2012. The positive polarity of the $B_{\text{los}}$ in the emission lines is again consistent with the presence of an accretion spot always visible and close to the pole. This geometry is similar to what has been found for other cTTs \cite[see e.g.,][]{2014MNRAS.437.3202J, 2020MNRAS.497.2142M}. 

We derived an estimate of the magnetic obliquity from the emission lines using the third equation of \cite{1967ApJ...150..547P}. This equation assumes a pure dipole, which is a simplification of the magnetic field present in the accretion shocks. 
It is the field present in these accretion shocks that is understood to best probe the global stellar magnetic field that reaches the circumstellar disk. 
The magnetic obliquities derived from the accretion-powered emission lines are therefore a reasonable reflection of the real dipole present above the surface of the star. 
For the He{\sc i} emission line, we find a magnetic obliquity of 26\si{\degree} for the 2010 epoch, and of 21\si{\degree} for the 2012 epoch. For the Ca{\sc ii} IRT, we find a magnetic obliquity of 16\si{\degree} for the 2010 epoch, and of 23\si{\degree} for the 2012 epoch. These estimates are consistent with the magnetic obliquity of 18\si{\degree} (+8)(-7) in 2011 derived by \cite{2020MNRAS.497.2142M}, using the He{\sc i} emission line and assuming one accretion spot. 

We also estimated the truncation radius using the values of $B_{\text{los}}$ in the emission lines, and find 
$r_{\text{trunc}}$ $\sim$ (5.2 $\pm$ 1.1) $R_\star$ for the He{\sc i} emission line in 2010, 
$r_{\text{trunc}}$ $\sim$ (3.9 $\pm$ 0.8) $R_\star$ for the Ca{\sc ii} IRT in 2010, 
$r_{\text{trunc}}$ $\sim$ (4.7 $\pm$ 1.2) $R_\star$ for the He{\sc i} emission line in 2012, 
and $r_{\text{trunc}}$ $\sim$ (3.9 $\pm$ 1.0) $R_\star$ for the Ca{\sc ii} IRT in 2012. 
We calculated the co-rotation radius as well, and find $r_{\text{co-rot}}$ = 6.1\,$R_\star$. We find that the truncation radius values are consistent with the co-rotation radius within the error bars. 

In conclusion, 
we find that DK\,Tau, presenting with significant veiling, has similar magnetic properties to the more moderately accreting cTTs studied so far. 
In addition, we find that DK\,Tau's outer disk axis is likely to be misaligned compared to its rotation axis by 38\si{\degree}. This poses questions with regards to standard models of circumstellar disk formation. More observations of cTTs are needed to better understand the prevalence of such misalignments, while the geometry of DK\,Tau's system requires additional studies to characterize it further.


\section*{Acknowledgements}

The authors thank A. Natta for helpful discussions and C. Delvaux for the sketch of DK\,Tau. We also thank the referee for valuable comments that improved the paper. 

Based on observations obtained at the Canada-France-Hawaii Telescope (CFHT) which is operated by the National Research Council of Canada, the Institut National des Sciences de l'Univers of the Centre National de la Recherche Scientifique of France, and the University of Hawaii. 
Based on observations obtained at the Télescope Bernard Lyot (TBL) which is operated by the Institut National des Sciences de l'Univers of the Centre National de la Recherche Scientifique of France. 

This work has made use of the VALD database, operated at Uppsala University, the Institute of Astronomy RAS in Moscow, and the University of Vienna.

This project has received funding from the European Research Council (ERC) under the European Union’s Horizon 2020 research and innovation programme under grant agreement No 743029 (EASY: Ejection Accretion Structures in YSOs), as well as grant agreement No 817540 (ASTROFLOW), grant agreement No 742095 (SPIDI) and grant agreement No 740651 (NewWorlds).

\bibliographystyle{aa}
\bibliography{references}


\clearpage
\newpage

\begin{appendix}

\onecolumn

\section{Stellar parameters} \label{ParamSiess}

\begin{figure*}[hbtp]
\centering
\includegraphics[scale=0.90]{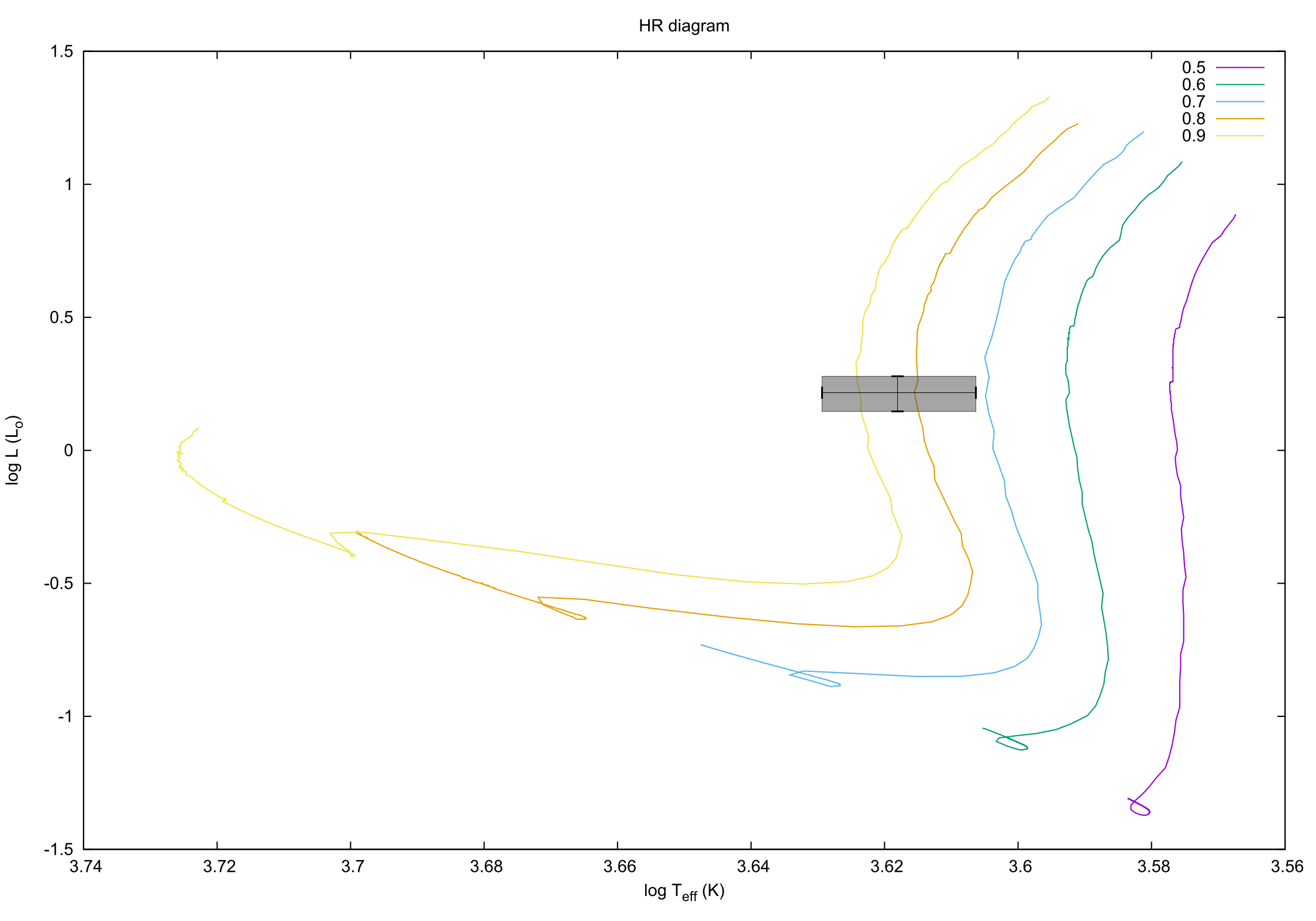}
\caption{Hertzsprung-Russell diagram with PMS evolutionary tracks from \cite{2000A&A...358..593S}. The colored lines correspond to different masses (in $M_\odot$). The gray rectangle highlights our values of $T_{\textrm{eff}}$ and $L_\star$ with their error bars. \label{SiessPlot}}
\end{figure*}

Using the \cite{2000A&A...358..593S} 
models\footnote{\url{http://www.astro.ulb.ac.be/~siess/pmwiki/pmwiki.php?n=WWWTools.PMS}}, 
we checked the compatibility of the values of $M_\star$, $T_{\textrm{eff}}$ and $L_\star$ and obtained Fig.\,\ref{SiessPlot}. The range of $T_{\textrm{eff}}$ and $L_\star$ (accounting for their error bars) that we derive correspond to masses that are close within 2$\sigma$ to the value of  $M_\star$ = 0.7\,$M_\odot$ quoted by \cite{2007ApJ...664..975J}. 

\FloatBarrier


\section{Photospheric absorption lines} \label{LSDprof}

\begin{figure*}[hbtp]
\centering
\includegraphics[scale=0.45]{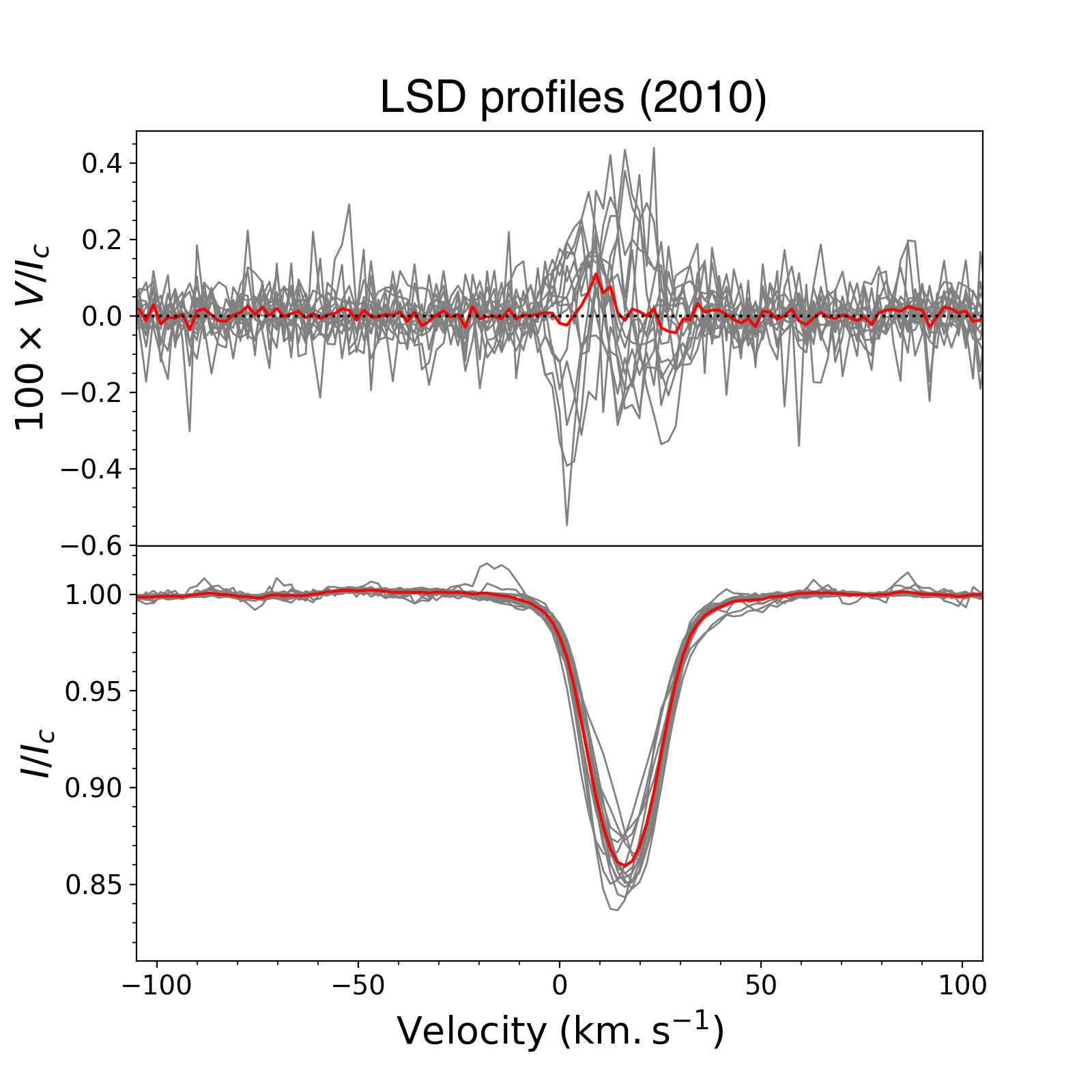}
\includegraphics[scale=0.45]{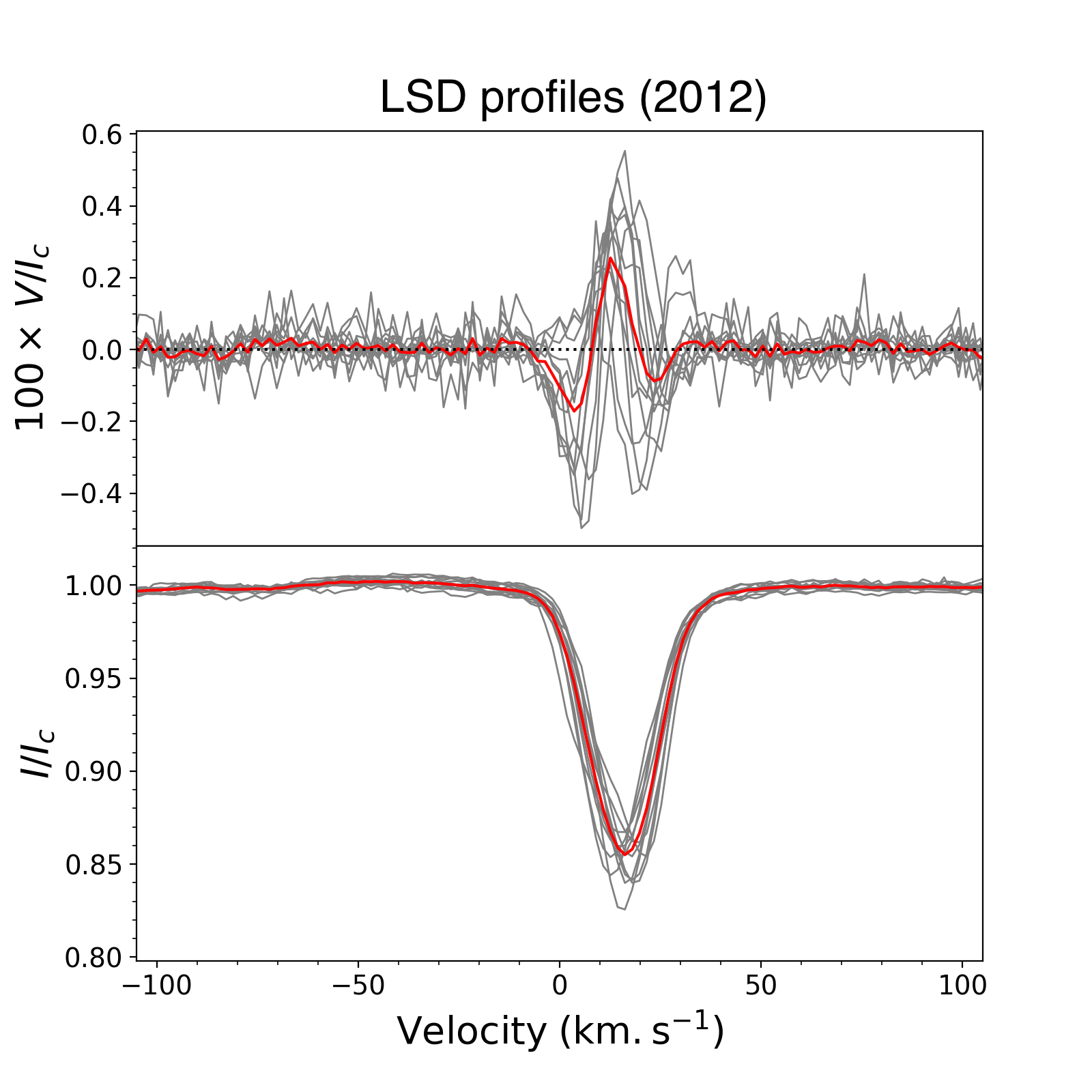}
\caption{Stokes\,V and Stokes\,I profiles (in gray) and average (in red) of the absorption lines, for the 2010 (left panels) and 2012 (right panels) observations. \label{LSDproffig}}
\end{figure*}

Figure\,\ref{LSDproffig} shows the Stokes\,V and Stokes\,I profiles of the photospheric absorption lines for both epochs. Table\,\ref{TableBlosAL} lists the values of the $B_{\text{los}}$, the line-of-sight magnetic field integrated over the visible hemisphere, for the photospheric absorption lines. 

\FloatBarrier

\begin{table}
\begin{center}
\caption{$B_{\text{los}}$ for the photospheric absorption lines. \label{TableBlosAL}}
\begin{tabular}{c c c}
\hline
Date & Rotation cycle & $B_{\text{los}}$ \\
(yyyy-mm-dd) & (8.2 day period) & (G) \\
\hline 
  2010-11-26 & 0.00 & -148.49 \\  
  2010-12-09 & 1.58 & 131.92 \\  
  2010-12-10 & 1.70 & 16.08 \\  
  2010-12-13 & 2.07 & -185.01 \\  
  2010-12-14 & 2.14 & -186.52 \\  
  2010-12-15 & 2.25 & -90.83 \\  
  2010-12-16 & 2.37 & 30.36 \\  
  2010-12-17 & 2.49 & 205.23 \\  
  2010-12-18 & 2.61 & 87.10 \\  
  2010-12-19 & 2.73 & 31.08 \\  
  2010-12-19 & 2.80 & 6.72 \\  
  2010-12-24 & 3.35 & 10.11 \\  
  2010-12-26 & 3.61 & 100.33 \\  
  2010-12-30 & 4.09 & -148.73 \\  
  2011-01-03 & 4.64 & 93.22 \\  
\hline
  2012-11-19 & 0.00 & -29.55 \\  
  2012-11-25 & 0.78 & -124.88 \\  
  2012-11-28 & 1.15 & -17.76 \\  
  2012-11-29 & 1.28 & 81.99 \\  
  2012-12-01 & 1.52 & -15.75 \\  
  2012-12-02 & 1.63 & -78.84 \\  
  2012-12-04 & 1.89 & -74.67 \\  
  2012-12-07 & 2.24 & -5.17 \\  
  2012-12-09 & 2.55 & -87.57 \\  
  2012-12-10 & 2.60 & -102.34 \\  
  2012-12-12 & 2.81 & -35.25 \\  
  2012-12-23 & 4.18 & 26.34 \\  
\hline
\end{tabular}
\end{center}
\end{table}
\FloatBarrier


\section{Emission lines} \label{DecompPlots}

\begin{figure*}[hbtp]
\centering
\includegraphics[scale=0.55]{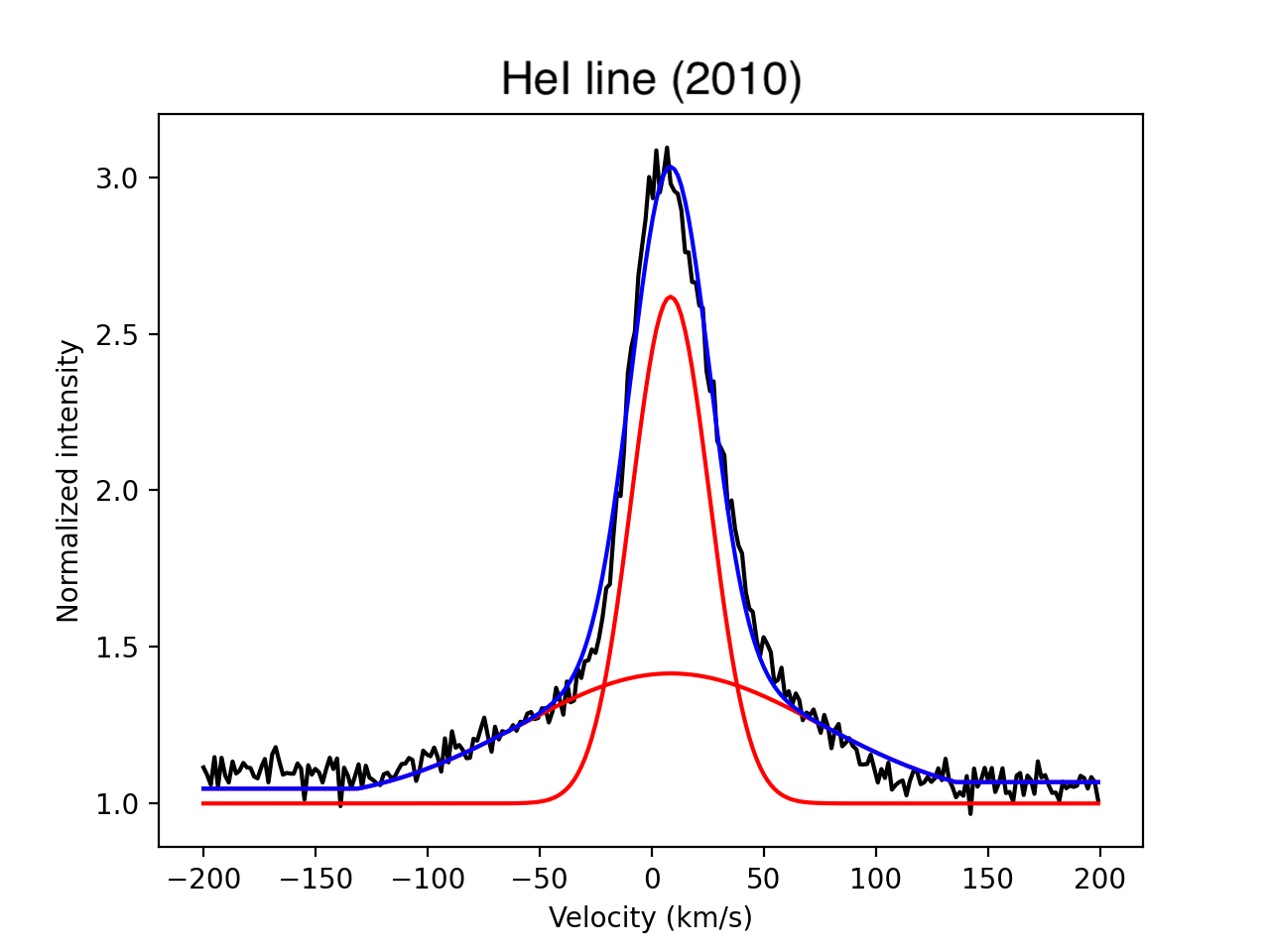}
\includegraphics[scale=0.55]{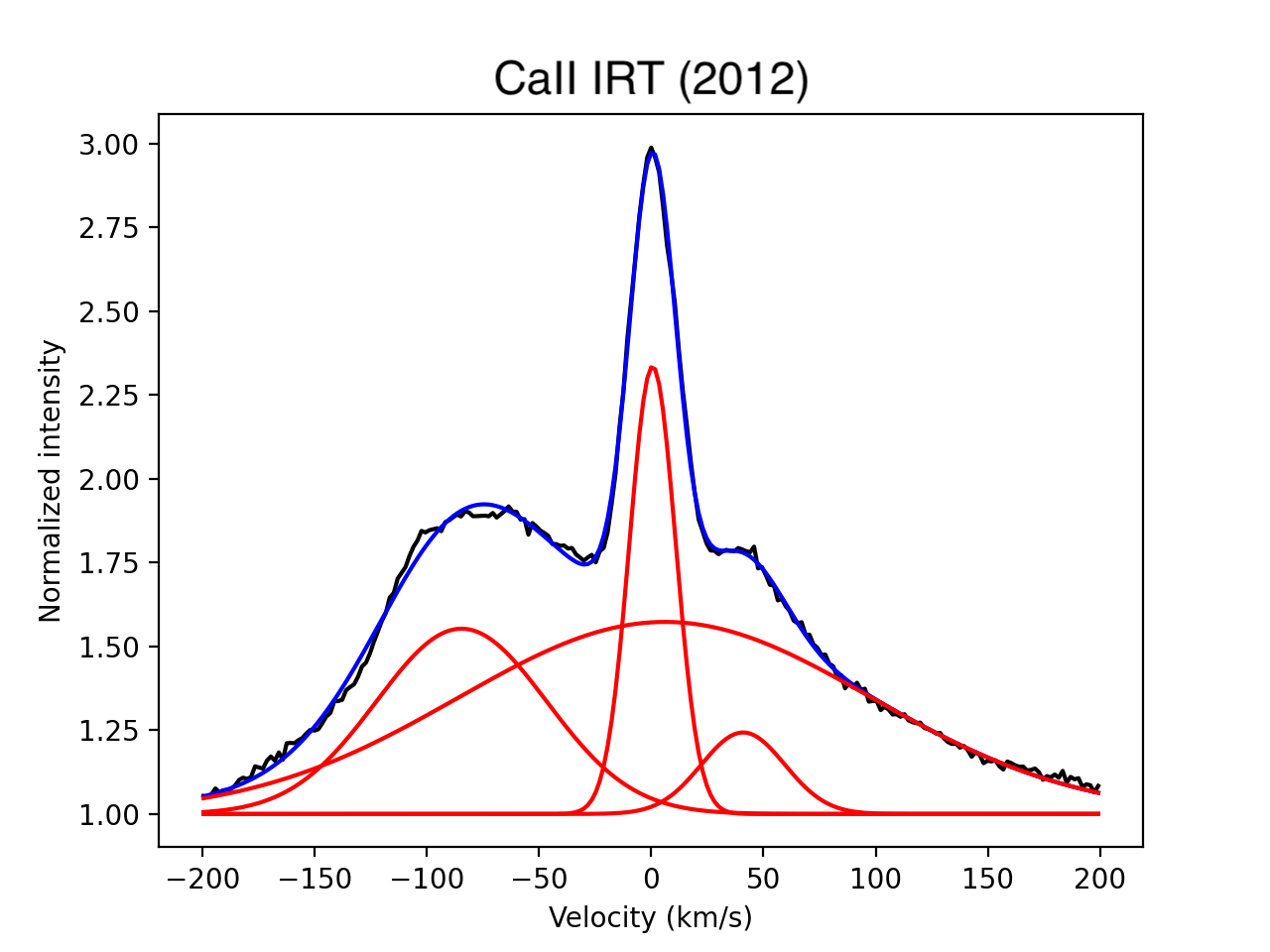}
\caption{Fit of the He{\sc i} line for the first night of the 2010 ESPaDOnS observations (left panel) and of the average of the Ca{\sc ii} IRT for the second night of the 2012 ESPaDOnS observations (right panel). The observed line is in black. The different components are in red and their sum is in blue. \label{LineFit}}
\end{figure*}

The emission lines associated with accretion shocks 
have multiple components (usually a broad and a narrow component). 
The narrow component is believed to come from the accretion shock, and that is the component we wish to isolate to probe the magnetic field in the shock region. 
We fit several components of each emission line that we studied and then subtracted all the components except the narrow one, to get a residual profile. Figure\,\ref{LineFit} shows two examples, for the He{\sc i} line (at 587.6\,nm) and for the average of the Ca{\sc ii} IRT (at 849.8\,nm, 854.2\,nm and 866.2\,nm). 

Table\,\ref{TableBlosEL} lists the values of the $B_{\text{los}}$, the line-of-sight magnetic field integrated over the visible hemisphere, for the He{\sc i} line and the Ca{\sc ii} IRT. 

Table\,\ref{TableEWHeI} lists the values of the equivalent width (EW) of the He{\sc i} line. 

\twocolumn

\begin{table}
\begin{center}
\caption{$B_{\text{los}}$ for the emission lines. \label{TableBlosEL}}
\begin{tabular}{c c c c}
\hline
Date & Rotation cycle & $B_{\text{los}}$ & $B_{\text{los}}$ \\
(yyyy-mm-dd) & (8.2 day period) & for He{\sc i} (G) & for Ca{\sc ii} (G) \\
\hline 
  2010-11-26 & 0.00 & 688.77 & 341.07 \\  
  2010-12-09 & 1.58 & 912.45 & 778.92 \\  
  2010-12-10 & 1.70 & 1601.15 & 781.50 \\  
  2010-12-13  & 2.07 & -201.64 & 450.74 \\  
  2010-12-14 & 2.14 & 970.06 & 610.67 \\  
  2010-12-15 & 2.25 & 1392.66 & 621.40 \\  
  2010-12-16 & 2.37 & 1473.35 & 949.06 \\  
  2010-12-17 & 2.49 & 1579.56 & 826.53 \\  
  2010-12-18 & 2.61 & 1321.08 & 699.57 \\  
  2010-12-19 & 2.73 & 1759.99 & 663.43 \\  
  2010-12-19 & 2.80 & 1023.12 & 760.05 \\  
  2010-12-24 & 3.35 & 1767.61 & 944.87 \\  
  2010-12-26 & 3.61 & 923.56 & 671.05 \\  
  2010-12-30 & 4.09 & 1198.59 & 426.48 \\  
  2011-01-03 & 4.64 & 803.45 & 771.46 \\  
\hline
  2012-11-19 & 0.00 & 1957.62 & 1411.28 \\  
  2012-11-25 & 0.78 & 1988.06 & 876.93 \\  
  2012-11-28 & 1.15 & 1130.63 & 838.51 \\  
  2012-11-29 & 1.28 & 939.35 & 782.20 \\  
  2012-12-01 & 1.52 & 479.01 & 300.70 \\  
  2012-12-02 & 1.63 & 606.14 & 397.38 \\  
  2012-12-04 & 1.89 & 1670.70 & 1052.93 \\  
  2012-12-07 & 2.24 & 1872.10 & 1149.95 \\  
  2012-12-09 & 2.55 & 1178.70 & 828.18 \\  
  2012-12-10 & 2.60 & 1349.05 & 561.12 \\  
  2012-12-12 & 2.81 & 1137.52 & 263.62 \\  
  2012-12-23 & 4.18 & 1435.56 & 850.50 \\  
\hline
\end{tabular}
\end{center}
\end{table}

\begin{table}
\begin{center}
\caption{EW of the He{\sc i} line. \label{TableEWHeI}}
\begin{tabular}{c c c}
\hline
Date & Rotation cycle & EW for \\
(yyyy-mm-dd) & (8.2 day period) & He{\sc i} ($\mathring{A}$)  \\
\hline 
  2010-11-26 & 0.00 & 1.37  \\  
  2010-12-09 & 1.58 & 1.49  \\  
  2010-12-10 & 1.70 & 1.54  \\  
  2010-12-13 & 2.07 & 1.32  \\  
  2010-12-14 & 2.14 & 2.44  \\  
  2010-12-15 & 2.25 & 2.04  \\  
  2010-12-16 & 2.37 & 2.05  \\  
  2010-12-17 & 2.49 & 1.46  \\  
  2010-12-18 & 2.61 & 1.74  \\  
  2010-12-19 & 2.73 & 1.50  \\  
  2010-12-19 & 2.80 & 0.86  \\  
  2010-12-24 & 3.35 & 2.20  \\  
  2010-12-26 & 3.61 & 1.50  \\  
  2010-12-30 & 4.09 & 1.95  \\  
  2011-01-03 & 4.64 & 1.43  \\  
\hline
  2012-11-19 & 0.00 & 2.61  \\  
  2012-11-25 & 0.78 & 1.74  \\  
  2012-11-28 & 1.15 & 1.61  \\  
  2012-11-29 & 1.28 & 1.44  \\  
  2012-12-01 & 1.52 & 1.43  \\  
  2012-12-02 & 1.63 & 1.78  \\  
  2012-12-04 & 1.89 & 2.57  \\  
  2012-12-07 & 2.24 & 2.12  \\  
  2012-12-09 & 2.55 & 1.91  \\  
  2012-12-10 & 2.60 & 1.67  \\  
  2012-12-12 & 2.81 & 1.19  \\  
  2012-12-23 & 4.18 & 1.91  \\  
\hline
\end{tabular}
\end{center}
\end{table}

\FloatBarrier

\onecolumn


\section{Bidimensional periodograms} \label{2DPer}

\begin{figure*}[hbtp]
\centering
\includegraphics[scale=0.31]{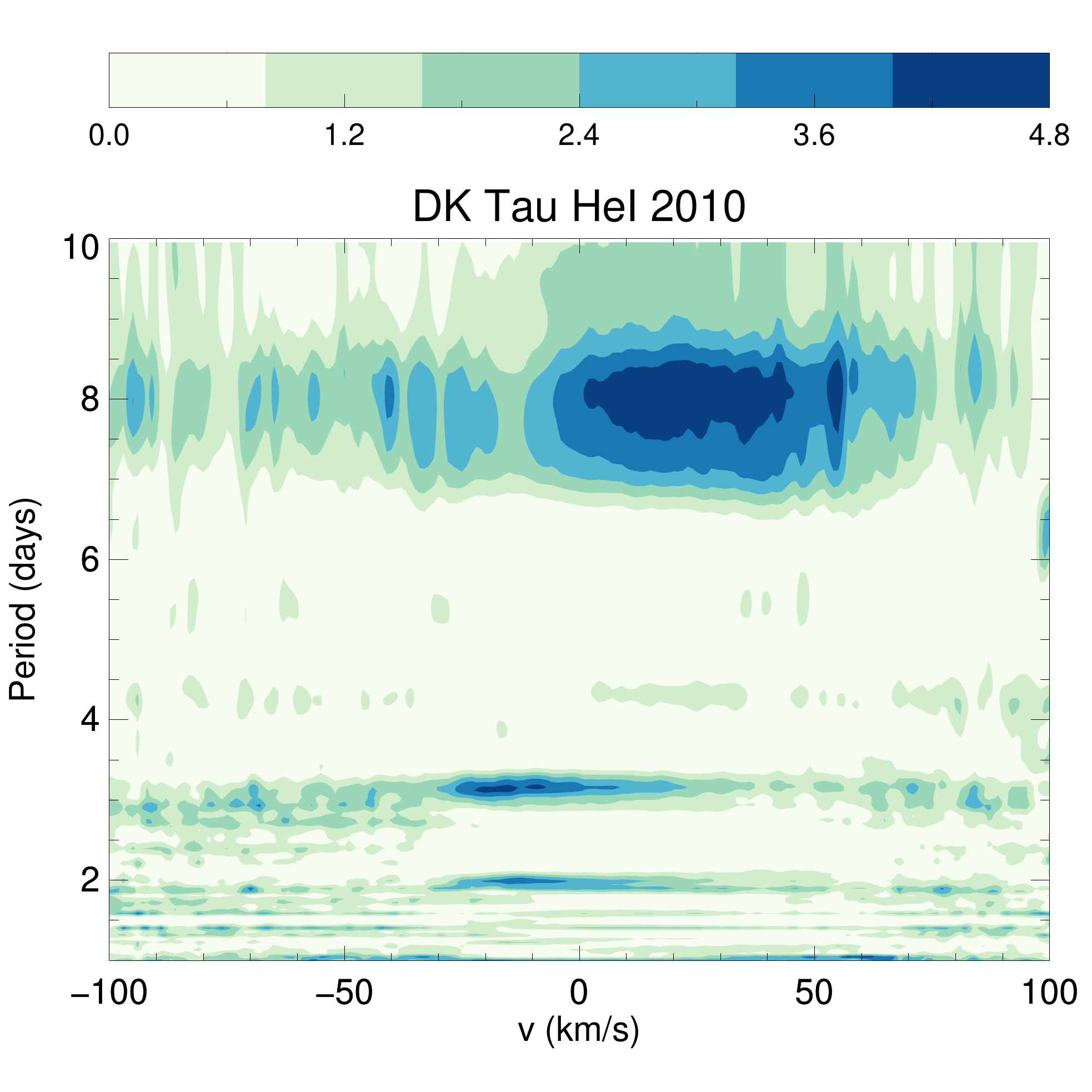}
\includegraphics[scale=0.31]{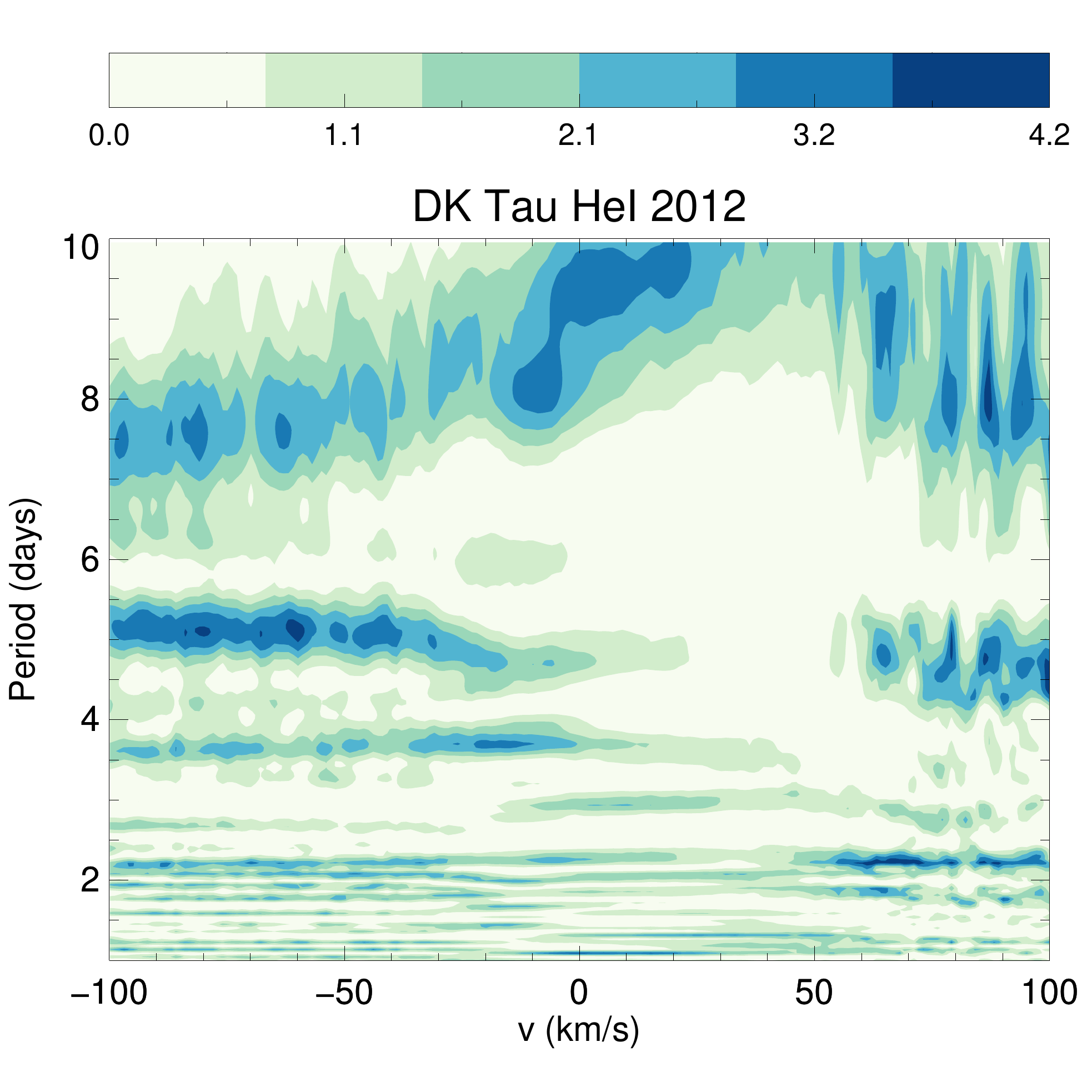}
\includegraphics[scale=0.31]{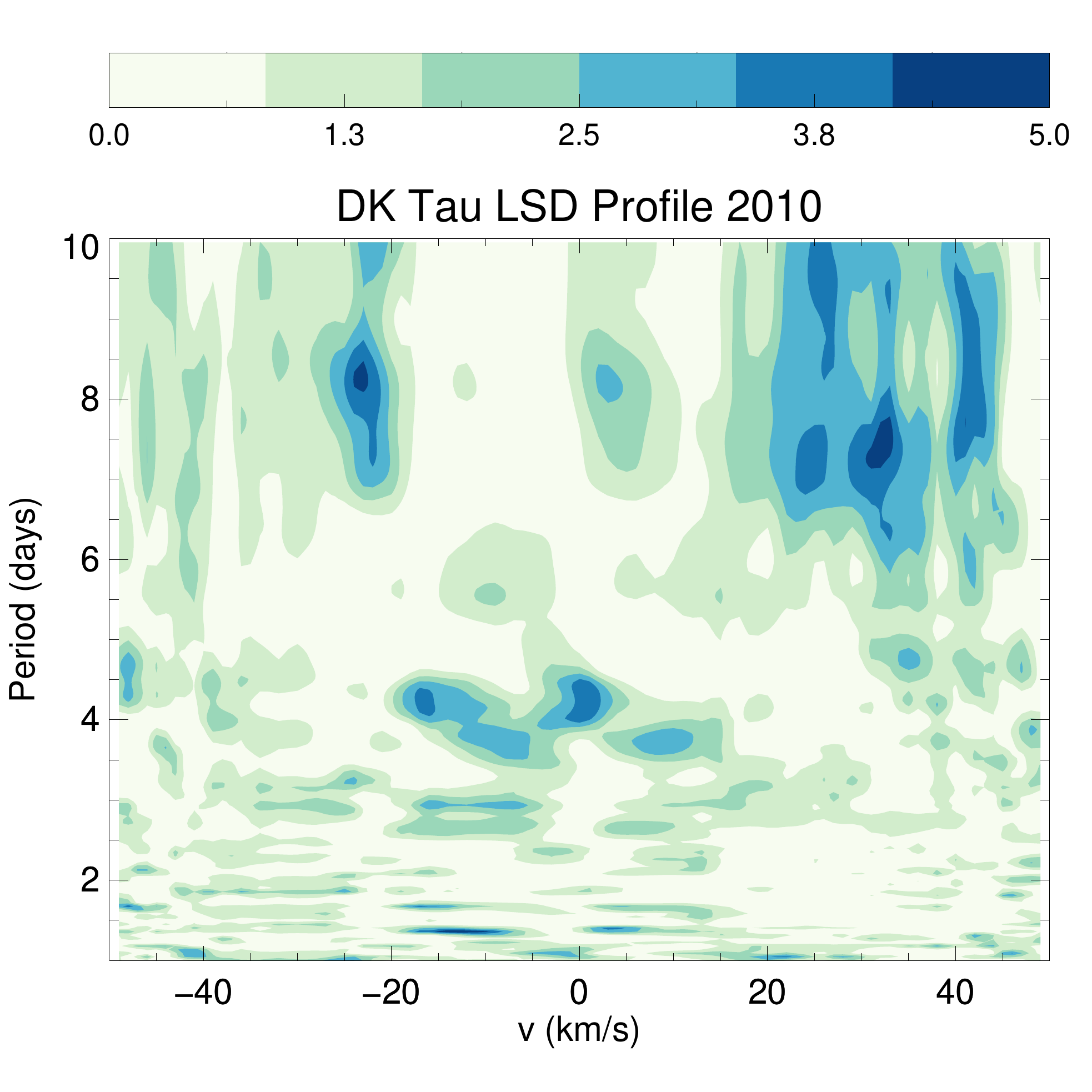}
\includegraphics[scale=0.31]{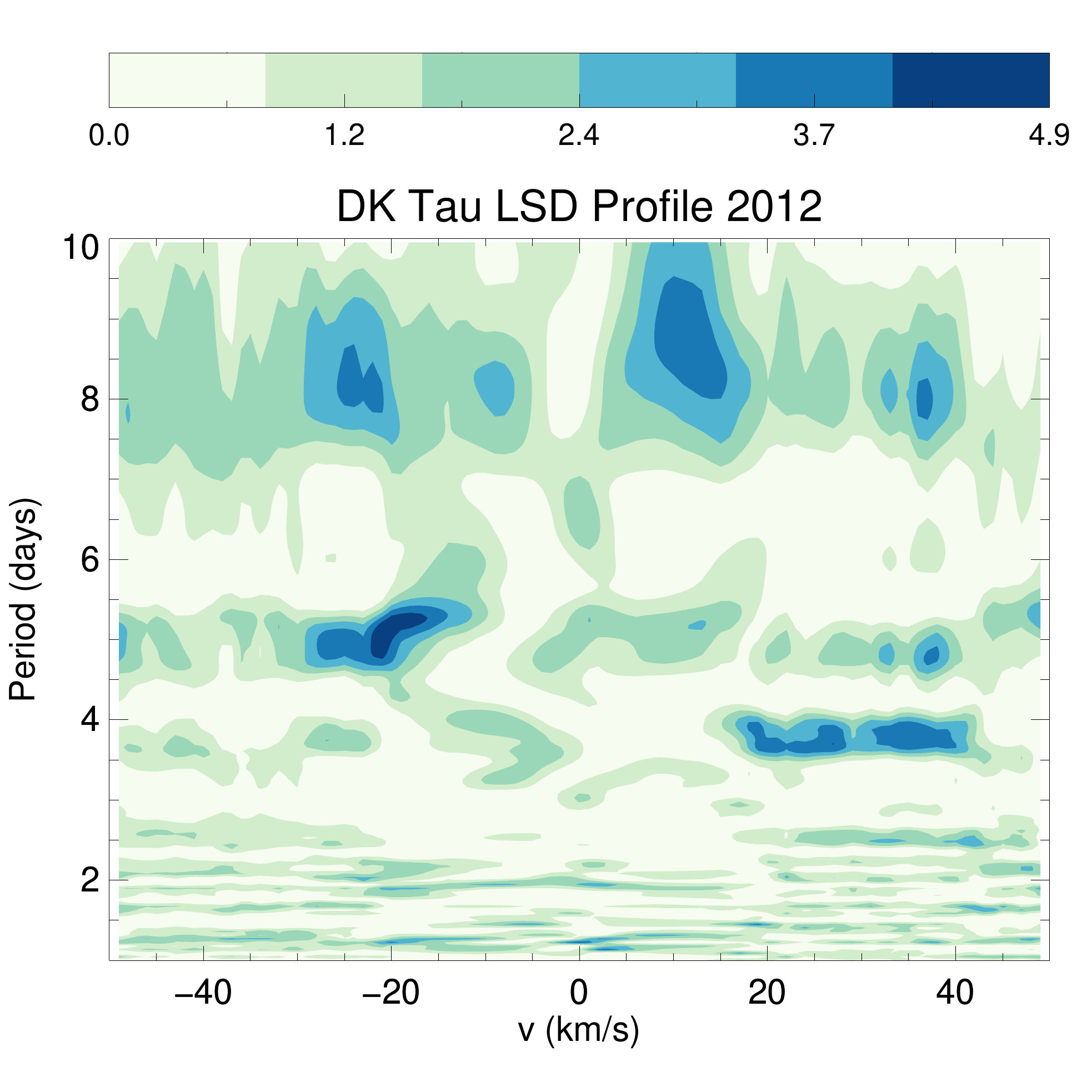}
\includegraphics[scale=0.31]{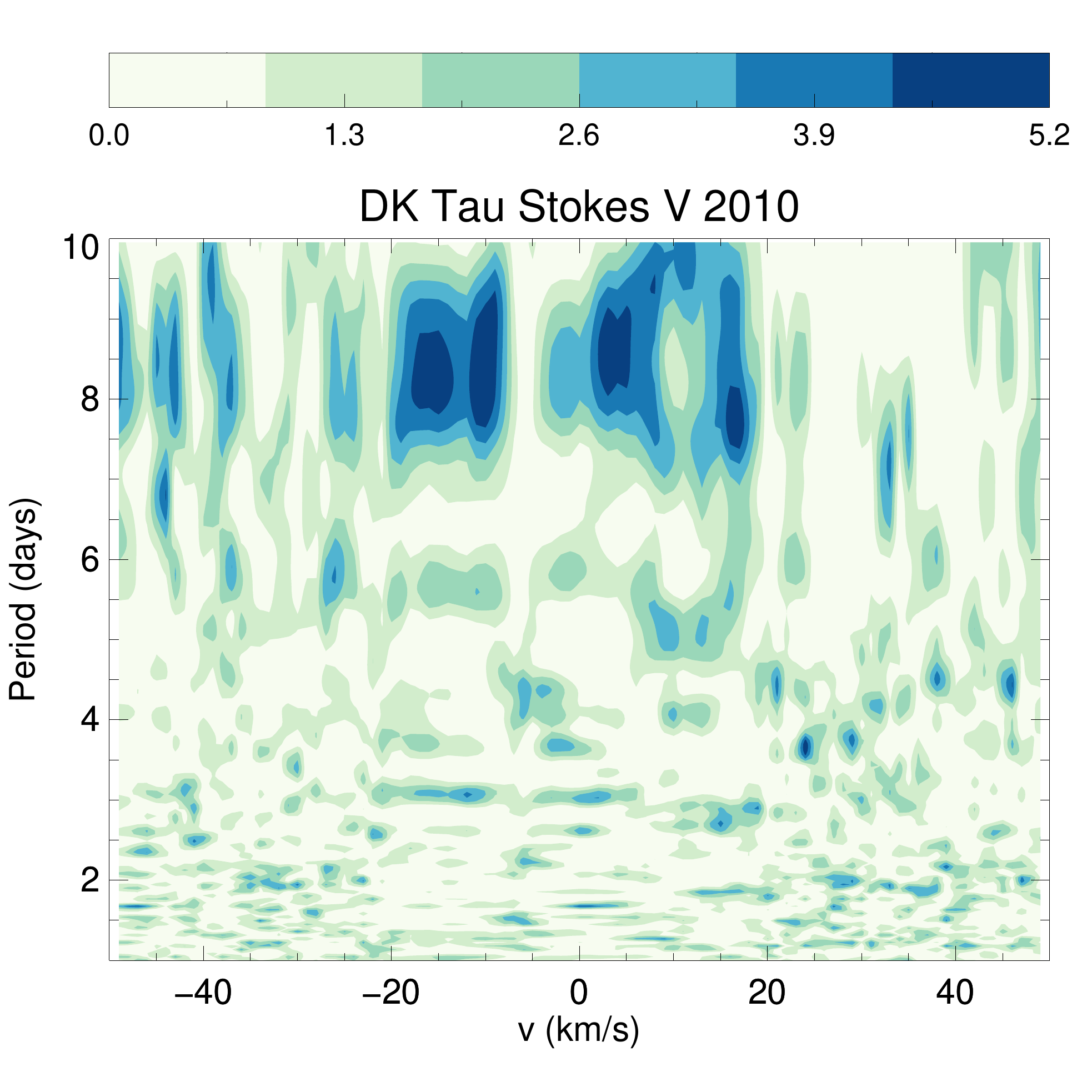}
\includegraphics[scale=0.31]{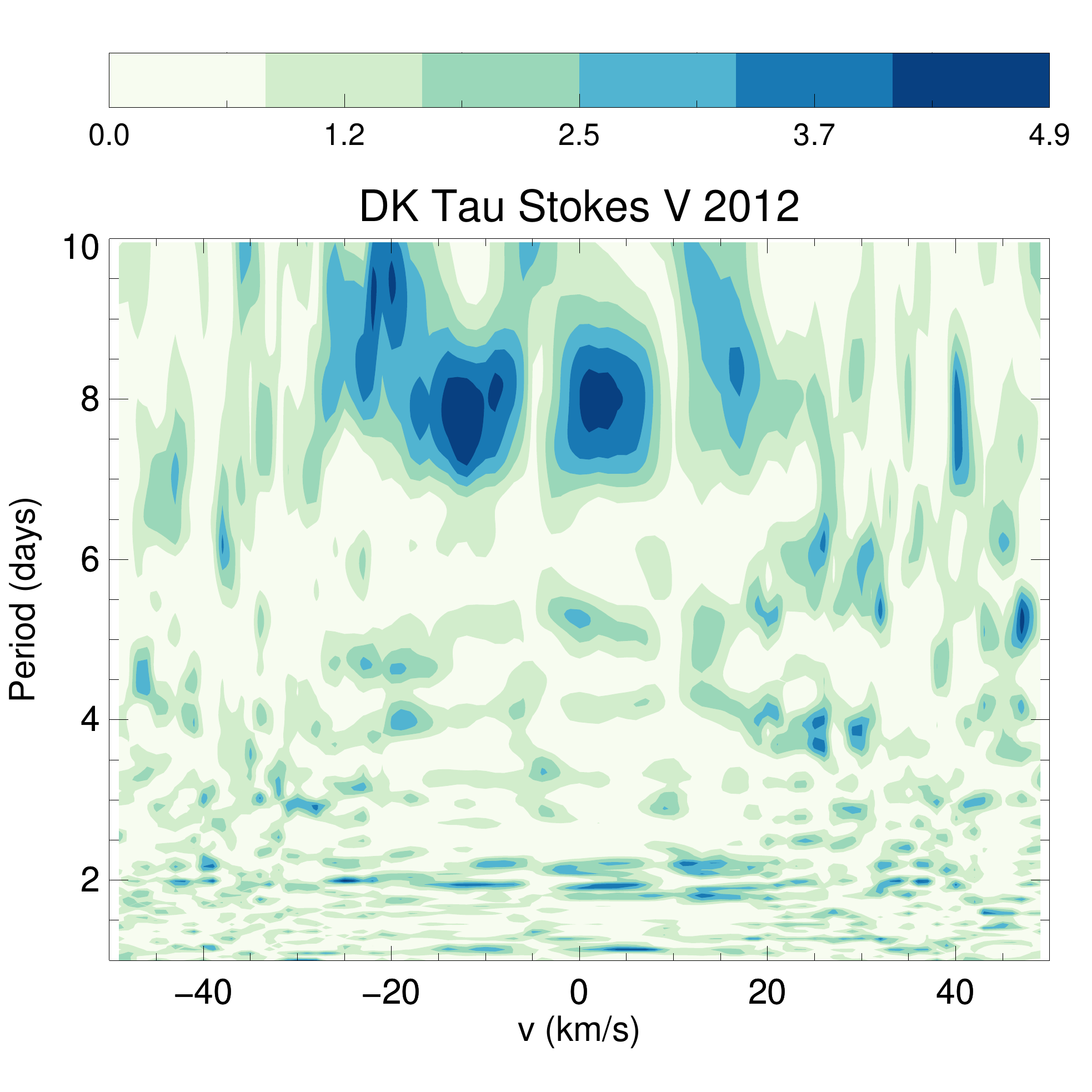}
\caption{Bidimensional periodograms of the intensity of the He{\sc i} (at 587.6\,nm) emission line (top panel), the Stokes\,I (middle panel) and Stokes\,V LSD profiles (bottom panel) of the photospheric absorption lines, for the 2010 (left panels) and 2012 epoch (right panels). 
The power of the periodogram is showed using the color code. A light color represents a zero power intensity, while a dark color represents the maximum power intensity.  \label{2DPerFig}}
\end{figure*}

Fig.\,\ref{2DPerFig} shows the bidimensional periodograms of the intensity of the He{\sc i} (at 587.6\,nm) emission line, the Stokes\,I and Stokes\,V LSD profiles of the photospheric absorption lines. Bidimensional periodograms analyze the intensity of the line in several bins over the velocity range. This allows us to see if different parts of the lines have different periods and therefore distinct origins. A dark color on the plots indicates a peak in the periodogram, meaning that a period was found, but a large spot indicates a large uncertainty.We would expect to find the same period in all the bins if the entire line has a single origin. This is seen for instance in the plot of the He{\sc i} line in 2010, where the same period is found for the whole redshifted part of the line. However, the width of the peak of the periodogram is very broad, indicating that there is a large uncertainty in this period.

\FloatBarrier


\section{H$\alpha$ lines} \label{Halpha}

\begin{figure}[hbtp]
\centering
\includegraphics[scale=0.55]{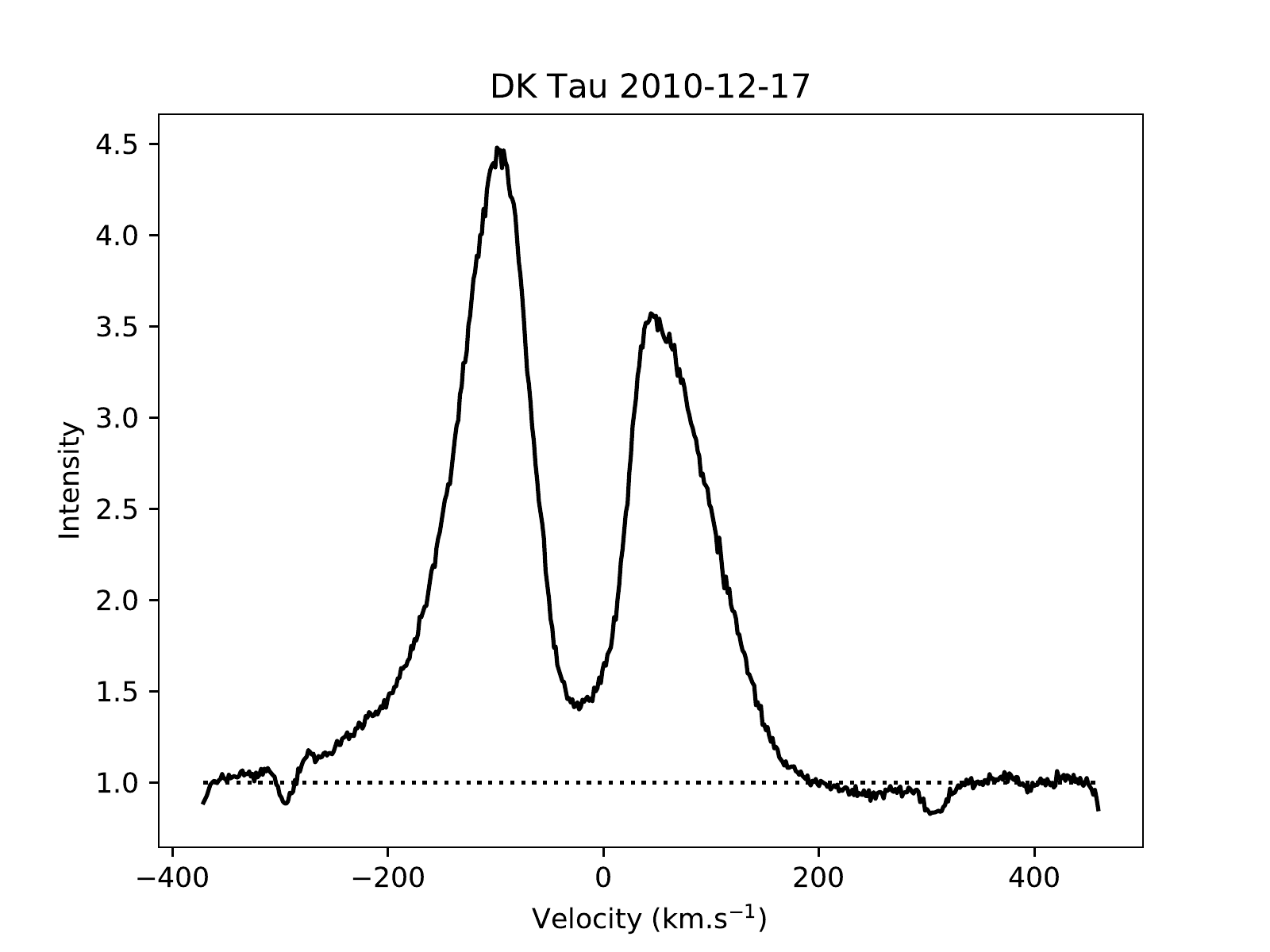}
\includegraphics[scale=0.55]{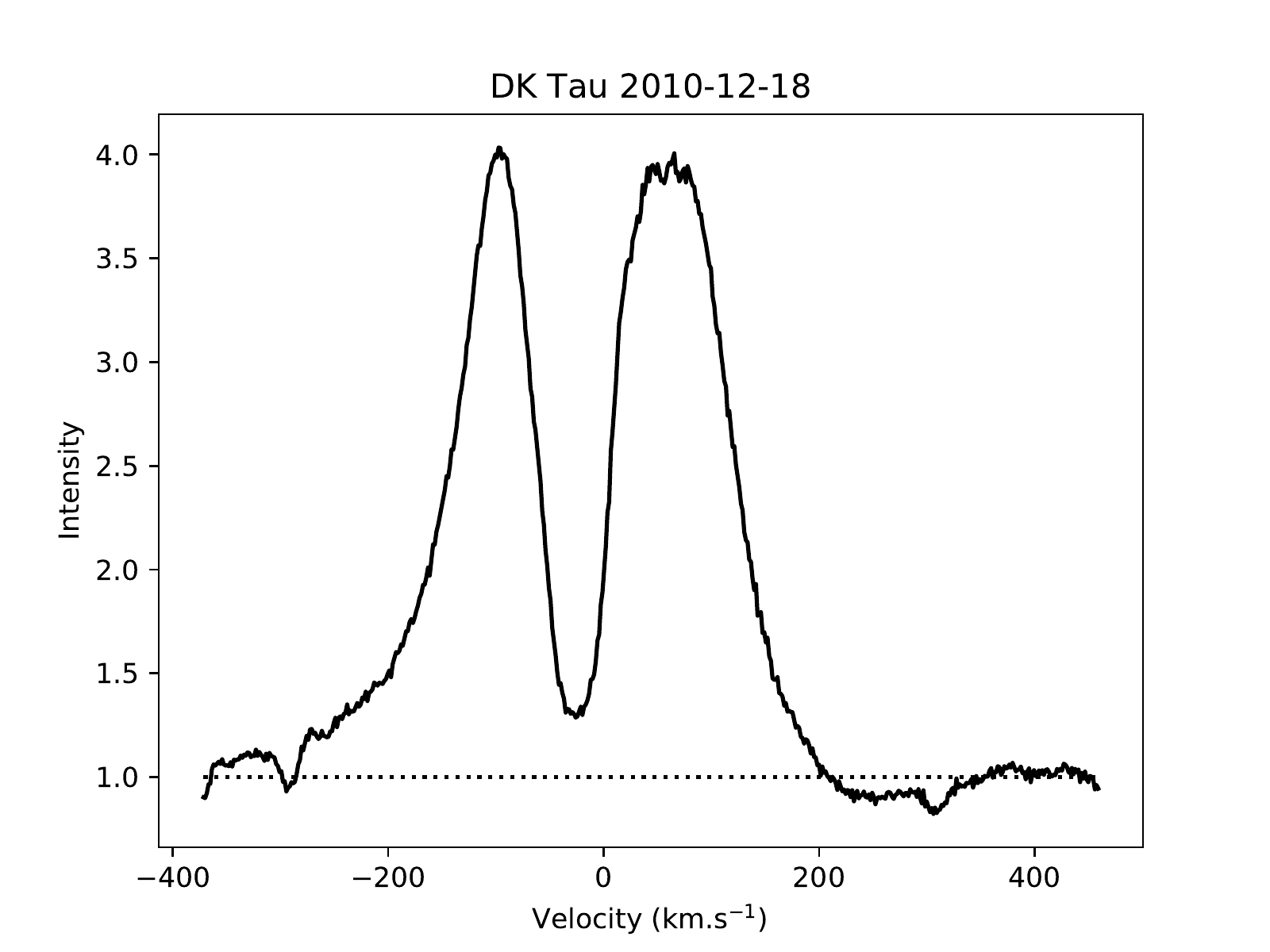}
\caption{H$\alpha$ line for the fourth night (right panel) and the fifth night (left panel) of the 2010 ESPaDOnS observations. The dotted line highlights the continuum. \label{HalpaFig}}
\end{figure}

Fig.\,\ref{HalpaFig} shows the H$\alpha$ line for the fourth and fifth night of the 2010 ESPaDOnS observations, corresponding to phase $\sim$0.5. For both nights, we see a small redshifted absorption, indicating that the accretion column is in our line-of-sight.  

\end{appendix}

\end{document}